\documentclass[twocolumn,prd,aps,tightenlines,floats,floatfix,preprintnumbers,nofootinbib,eqsecnum]{revtex4-2}%%

\def\sp{\kern +3pt}
\def\sm{\kern -3pt}
\def\spQ{\kern +6pt}

\def\bea{\begin{eqnarray}}
\def\eea{\end{eqnarray}}

\def\sfrac#1#2{{\textstyle \frac{#1}{#2}}}

\newcommand{\ket}[1]{|#1\rangle}

\def\be{\begin{equation}}
\def\ee{\end{equation}}
\def\ba{\begin{eqnarray}}
\def\ea{\end{eqnarray}}

\usepackage{graphics}
\usepackage{graphicx}
\usepackage{epsf}
\usepackage{amsmath}
\usepackage{amssymb}
\usepackage{xcolor}

\usepackage{ulem}

\begin{document}

\phantom{0}    
\vspace{-0.2in}  
\hspace{5.5in}

% include preprint number option
%  preprint number LFTC-24-06/89  
%\preprint{{\bf LFTC-24-06/89}}
%\preprint{{\bf Version S7}}

\vspace{-1in}%\parbox{1.5in}{ \vspace{-9.6in}}  % moves the preprint box down

\title
{\bf Weak interaction axial form factors of the octet baryons in nuclear medium}
\author{G.~Ramalho$^1$, 
K.~Tsushima$^2$ and Myung-Ki Cheoun$^1$}
\vspace{-0.1in}

\affiliation{$^1$Department of Physics and OMEG Institute, 
Soongsil University, \\
Seoul 06978, Republic of Korea
\vspace{-0.15in}}
\affiliation{$^2$Laborat\'orio de 
  F\'{i}sica Te\'orica e Computacional -- LFTC,
  Universidade Cidade de  S\~ao Paulo, 
01506-000,   S\~ao Paulo, SP, Brazil}

\vspace{0.2in}
\date{\today}

\phantom{0}

\begin{abstract}
We study the axial-vector and the induced pseudoscalar 
form factors associated with the weak transitions between the octet baryon members
in nuclear medium, using a covariant constituent quark model.
We extend previous calculations of the axial transition form
factors from the vacuum (free space) to the nuclear medium (symmetric nuclear matter).
The extension of the model to the nuclear medium takes into
account the modifications of the properties of hadrons in the medium
(masses and coupling constants),
as determined by the quark-meson coupling model.
The axial-vector ($G_A$) and the induced pseudoscalar ($G_P$) form factors
are evaluated for different values of the nuclear density $\rho$
in terms of the square transfer momentum $q^2= -Q^2$.
We conclude that, in general, the $G_A$ and $G_P$ form factors
are reduced in the nuclear medium.
The reduction is stronger for light baryons and high densities.
The medium modifications are milder for the 
heavier octet baryons, particularly at large $Q^2$.
The calculations presented here can be used to
estimate the cross sections of neutrino and antineutrino scattering
with nucleus, and neutrino and antineutrino scattering
with hyperons bound to a nucleus,
as well as those in the cores of compact stars.
\end{abstract}

%\phantom{0}
%\vspace{7.0in}
%\vspace{-6in}
\vspace*{0.9in}  % sets how far the title is below the preprint box
\maketitle

\section{Introduction \label{secIntro}}

In recent years a considerable effort 
has been made in the experimental and theoretical 
studies of the internal structure of hadrons,
including the electromagnetic and weak structures
of the nucleon and other baryons in the 
vacuum~\cite{NSTAR,Aznauryan12,Ramalho-Pena23,Spectator-Review,Bernard02,Gaillard84,Gorringe04,
Athar-Review22,NSTAR,AxialFF,Octet,Formaggio12a},
and in nuclear matter, and dense nuclear
medium~\cite{Rosa22a,Ankowski22a,Patino22a,Ruso22a,Hayano,Kienle,Meissner,Vogl,JLabbook,Pate24a}.
The normal nuclear matter may be characterized by
the saturation density $\rho_0 \simeq 0.15$ fm$^{-3}$,
and the energy per nucleon $\simeq -15.7$ MeV.
The structure of the hadrons is modified 
when immersed in the nuclear medium.
Studies in vacuum are important to set a reference
to study in nuclear matter.
Studies in the nuclear medium are important for
the understanding of environments with dense nuclear matter
from the high-energy nucleus-nucleus collisions
to the cores of compact stars
(neutron stars and
magnetars)~\cite{Octet2,QMCReview,Cheoun13a,Cheoun13b,Tsushima22a,QMCEMFFMedium,Miyatsu15a}.
At the moment, however, there are not many theoretical
studies about the internal structure modifications
of baryons in nuclear matter.

The present work highlights the study of
the axial structure of baryons in nuclear matter.
The main motivation is to develop a formalism that can be used to make
predictions of physical properties of baryons at relatively high densities
as well as moderate, and large $Q^2$,
based on the degrees of freedom manifest in the free space:
valence quarks and meson cloud excitations of the baryon cores.
Within the formalism one can estimate the modifications
of the axial transition form factors due to the nuclear medium,
by comparison with the corresponding functions in the free space~\cite{Octet2,Octet3}.
Deviations between the in-medium and in-vacuum form factors
are a manifestation of the role of the medium on the internal structure
of the baryons.
The possible effects were observed already in
electromagnetic reactions~\cite{JLabbook,Dieterich01,Paolone10,Strauch03},
although the interpretations of the observed modifications
in connection with the European Muon
Collaboration (EMC) effect~\cite{EuropeanMuon83a} are still in
debate~\cite{Xing23a,Kim24a}.
The calculations presented here can be used in
the study of reactions of neutrinos/antineutrinos
with nucleus and reactions of neutrinos/antineutrinos with hyperons
in dense nuclear matter.
The methods developed here for the octet baryons
can be extended in the future to the other baryon systems,
like decuplet baryons and 
transitions between octet baryons and decuplet baryons,
as well as for densities larger than the normal nuclear matter.

Our starting point is the axial structure of the octet baryons in free space,
in particular the $G_A$ and $G_P$ form factors for elastic transitions
and semileptonic decays (like the $n \to p + e^- + \bar{\nu}_e$ beta
decay)~\cite{Bernard02,Gaillard84,Gorringe04,Athar-Review22,AxialFF}.

Although there are not many recent developments from the experimental side,
there has been a significant progress in lattice QCD simulations for
weak axial-vector couplings and transition form factors of the
nucleon~\cite{Meyer22a,Chang18a,Alexandrou11a,Alexandrou24a,Jang20-24,Djukanovic22a,ARehim15a,Hasan19a,Tomalak23a},
as well as for the axial-vector couplings of the octet
baryons~\cite{Lin09,Erkol,Bali23a}.
Lattice QCD simulations for pion mass values close to the physical
limit reproduce accurately the experimental
value of the nucleon axial-vector coupling, $g_A = G_A(0)$, but
there are still discrepancies on the axial-vector
form factors $G_A(Q^2)$ for finite $Q^2$~\cite{Meyer22a,Chang18a}.

The combined analysis of lattice simulations and experimental
data for large $Q^2$, 
with model calculations based on different frameworks, 
suggests that the octet baryon weak axial-vector form factors
can be interpreted as the result of a dominant contribution 
associated with the valence quark and a component associated
with the meson cloud dressing of the baryon cores~\cite{AxialFF,GAholography}.

In a previous work we have calculated the weak axial 
transition form factors between
the octet baryon members in the vacuum~\cite{AxialFF}.
The model in the vacuum was derived from the framework of the 
covariant spectator quark model~\cite{Spectator-Review,AxialFF,Nucleon,Nucleon2,Omega},
developed in a first stage for the study of the electromagnetic structure 
of the baryons, including the nucleon
and nucleon resonances~\cite{Spectator-Review,NDelta,NDeltaTL,Lattice,Roper,N1535,N1520}.

The model based on the SU(6) spin-flavor symmetry for
the quark states has been applied in particular,
in the study of the octet baryons, the decuplet baryons
and transitions between the octet and decuplet
baryons~\cite{Octet,Octet2,Octet3,Sigma0Lambda,Omega2,OctetDecuplet1,DecupletDecays,OctetDecupletD1,OctetDecupletD2,HyperonFF}.
The starting point is the representation of the quark weak axial structure in terms of
quark axial form factors parametrized in a vector meson dominance form.
The free parameters of the models are fixed by the study 
of the weak axial form factors ($G_A$ and $G_P$) in the lattice QCD regime
for the nucleon.
Once the model is calibrated (quark form factors and radial wave functions),
the calculations are extrapolated for the octet baryons in the physical regime.
The contributions of the valence quarks are combined with
a phenomenological SU(3) parametrization of the meson cloud contribution.
The comparison between model estimates for the nucleon and
the experimental data is used to estimate the magnitude
of the contributions from the meson cloud effects.

The model for the vacuum is in the present work extended to
the study of the octet baryon weak axial form factors
in symmetric nuclear matter.
The methodology used is based on the extension of a model developed previously
in vacuum~\cite{AxialFF} to the nuclear medium. 
The extension uses the quark-meson coupling (QMC) model~\cite{QMCReview},
to estimate the masses of the baryons and the mesons,
as well as the baryon-meson couplings in nuclear medium.
The QMC model extends the properties of the
bag model~\cite{QMCReview,Thomas84,Lu01a}
to the nuclear matter.
The cloudy bag model has already been combined
with the covariant spectator quark model in the study
of the electromagnetic transitions
between baryon states in vacuum~\cite{OctetDecupletD2,OctetDecupletD1,DecupletDecays}.
In these works the correspondence between the
quark effective electromagnetic structure in the two models 
is used to calculate the electromagnetic interaction with baryons in intermediate states.
The valence quark component of the model in medium
is obtained considering the natural generalization of the covariant
spectator quark model based on the vector meson dominance
parametrization of the quark current and radial wave functions
expressed in terms of the mass of the baryons~\cite{Octet2}.
The extension of the meson cloud component is performed
using an effective SU(3) baryon-meson parametrization~\cite{AxialFF} and 
taking into account the medium modifications of the baryon-meson couplings.
The weak axial form factors are then calculated in nuclear medium
for different values of the nuclear density $\rho$.

The first studies of the electromagnetic and
axial structure of baryons were made
using Skyrme and soliton models~\cite{Skyrme61a,Adkins83a,Nyman87a,Meissner87a,Meissner89a,Alberto88a,Meier97a,Silva05a,Ledwig14a,Yang15,Christov95a,Rakhimov98a,Yakhshiev03,Smith04,Yakhshiev14a}
and the QMC model~\cite{QMCReview,QMCEMFFMedium,Tsushima22a,Cheoun13a,Cheoun13b}. 
Other studies of the axial structure of the nucleon and octet baryons
based on different frameworks in vacuum
can be found in
Refs.~\cite{Kubodera80s,Cheng22a,Liu22a,Eichmann12a,Chen21a,Chen22b,Kirchbach92a,Merten02a,Boffi02a,BCano03,JDiaz04a,Khosonthongkee04a,Pasquini07a,Adamuscin08a,Liu14a,Dahiya14a},
while, studies of the electromagnetic and
axial structure of the nucleon and octet baryons 
in medium can be found in
Refs.~\cite{Cai23a,Atayev22a,Dominguez23a,Noro24a,Hen17a,Lu99a,Miller02a,Horikawa05,Cloet09a,deAraujo18,Vega21a,Frank96,Singh18a,Osterfeld92a,Saito24a}.

This article is organized as follows:
In the next section we define the currents and
transition form factors associated with the weak transitions
between the octet baryon states.
In Sec.~\ref{secVacuum} we review the covariant spectator
formalism for the octet baryon states and the weak axial interactions
in the vacuum.
The extension of the model to nuclear matter is discussed
in Sec.~\ref{secMedium}.
Numerical results for the weak axial form factors $G_A$ and $G_P$ in
the nuclear medium are presented in Sec.~\ref{secResults},
and they are also compared with those in free space.
In Sec.~\ref{secResultsCS} we exemplify how
the model calculations can be used
for the study of neutrino/antineutrino scattering with nucleus.
The outlook and conclusions are given in Sec.~\ref{secConclusions}.

\section{Weak Interaction Axial form factors
\label{secFormalism}}

\subsection{Axial-vector current}

The weak axial form factors ($G_A$ and $G_P$)
for the transitions between 
the octet baryon members $B$ and $B'$ can be represented as
\ba
J_5^\mu =
\frac{1}{2}\bar u_{B'}(P_f) \left[
G_A(Q^2) \gamma^\mu + G_P \frac{q^\mu}{2 M_{BB'}} 
\right]\gamma_5 u_B(P_i), \nonumber \\
\label{eqJA1}
\ea
where $u_{B'}$ and $u_B$ are the corresponding Dirac spinors,
and $M_{BB'}$ is the average mass of the final (mass $M_{B'}$)
and initial (mass $M_B$) baryons $M_{BB'} = \frac{1}{2}(M_{B'} + M_B)$.
The factor $1/2$ in Eq.~(\ref{eqJA1}) is included to
be consistent with the nucleon case~\cite{AxialFF}.
The form factors $G_A$ and $G_P$ depend then on the indices $B$ and $B'$.
For simplicity, we omit those indices when possible.

The expression Eq.~(\ref{eqJA1}) is obtained after a proper projection in
the flavor space.
Those flavor operators can be represented
in terms of the Gell-Mann matrices $\lambda_i$ ($i=1,..,8$)~\cite{Gaillard84}
as discussed in Appendix~\ref{appGM-matrices}.  
For the present study the cases of interest
are the operators $I_0= \lambda_3$ 
(neutral transitions), 
$I_\pm$ (increases/decreases the isospin projection)
and $V_\pm$ (decreases/increases the number of strange quarks).

The form factor $G_P$ is suppressed in the neutrino-nucleon
cross sections by a factor $(m_\ell/M_{BB'})^2$, where $m_\ell$
is the lepton mass, the counterpart of the neutrino
(electron or muon)~\cite{Cai23a}.
One expects then that the effect of $G_P$ can be neglected
in $\nu_e N$ and $\bar \nu_e N$ reactions.
Nevertheless, in the present work,
we calculate the function $G_P$ for several reasons:
(i) The theoretical effort is equivalent to the calculation of $G_A$,
(ii) it may be necessary for calculations of $\nu_\mu N$
and $\bar \nu_\mu N$ reactions, and
(iii) there is the chance that the function $G_P$
may be relevant for other observables or
physical processes to be proposed in a near future.
(see e.g., Ref.~\cite{Gorringe04}).

\subsection{Experimental status}
\label{secPhenomenology}

We start with the case of the nucleon.
The function $G_A$ can be measured by 
neutrino scattering and pion electroproduction off nucleons.
Both experiments suggest a dipole-type function dependence
$G_A(Q^2)= G_A(0)/(1+ Q^2/M_A^2)^2$, 
where the value of $M_A$ varies between 1.03 and 1.07 GeV
depending on the method~\cite{Bernard02,Schindler07a}.

To represent the experimental data in a general form 
we consider the interval between the two functions, $G_A^{{\rm exp}-}$
and $G_A^{{\rm exp}+}$, given by~\cite{AxialFF}
\ba
G_A^{{\rm exp}\pm}(Q^2)= \frac{G_A^0 (1 \pm \delta)}{
\left( 1 + \frac{Q^2}{M_{A\pm}^2} \right)^2},
\label{eqGAexp}
\ea
where $G_A^0=1.2723$ is the nucleon experimental value of $G_A(0)$~\cite{PDG2022},
$\delta =0.03$ is a parameter that expresses the precision 
of the data, and $M_{A-}=1.0$ GeV and  $M_{A+}=1.1$ GeV are,
respectively, the lower and upper limits for $M_A$.
The central value of the parametrization~(\ref{eqGAexp})
can be approximated by the dipole function form with $M_A \simeq 1.05$ GeV.
This value is almost identical to the
recent analysis from antineutrino-proton scattering~\cite{Cai23a}.

Most of the data analyses  are restricted to the region
$Q^2 < 1$ GeV$^2$~\cite{Bernard02}.
Recently the nucleon axial-vector $G_A$ form factor
was determined in the range $Q^2=2$--4 GeV$^2$
at CLAS/JLab~\cite{Park12}.
The new data are consistent with the parametrization~(\ref{eqGAexp}).

There are also nucleon data for $G_P$ in the range $Q^2=$0--0.2 GeV$^2$ 
determined by muon capture~\cite{Bernard02} and
from pion electroproduction data using the low-energy theorem~\cite{Choi93}.

Concerning the other members of the baryon octet,
there are data for $G_A(0)$ determined by the semileptonic decays 
indicated in Table~\ref{tabDataGA}.
The possible transitions between the octet baryon members are
discussed in the next section.

\begin{table}[t]
\begin{center}
\begin{tabular}{l c}
\hline
\hline
Transition  &   $G_A(0)$ \\
\hline
 $n \to p$  & $1.2723 \pm 0.0023$ \\
 $\Lambda \to p$ &    	$-0.879 \pm 0.018$         \\
 $\Sigma^- \to n$ &      $0.340 \pm 0.017$ \\
 $\Xi^- \to \Lambda$ &          $-0.306 \pm 0.061$  \\
 $\Xi^0 \to \Sigma^+$ &   	$1.22 \pm 0.05$  \\
\hline
\hline
\end{tabular}
\end{center}
\caption{\footnotesize{Experimental values for $G_A(0)$~\cite{PDG2022}.}}
\label{tabDataGA}
\end{table}

We discuss next, how the axial form factors
can be estimated in the context of a quark model 
with meson cloud dressing of the valence quark core.

\subsection{Theory} 
\label{secTheory}

The theoretical calculation of the transition form factors 
within a quark model framework starts with 
the representation of the physical baryon state 
in the form~\cite{AxialFF} 
\ba
\left|B \right>
= \sqrt{Z_B} \left[ \left| qqq \right> 
+ c_B \left| {\rm MC}\right> \right],  
\label{eqBaryonS}
\ea
where $\left|qqq\right>$ is the three-quark state
and $c_B \left| {\rm MC}\right>$ represents the baryon-meson
state associated with the meson cloud dressing.
The coefficient $c_B$ is determined by the normalization 
$Z_B (1+ c_B^2)=1$,
assuming that  $\left| {\rm MC}\right>$ is  normalized 
to unity.

In this representation $Z_B= \sqrt{Z_B} \sqrt{Z_B}$
{\it measures} the probability of 
finding the $qqq$ state in the physical baryon state.
Consequently, $1-Z_B$ {\it measures} 
the probability of finding the dressed baryon
(simultaneously the $qqq$ plus meson cloud) component
in the physical state.

In Eq.~(\ref{eqBaryonS}), we include   
only the first order, or ''one meson in the air'' correction for the meson cloud,
associated with the baryon-meson states.
In principle, we should also include corrections associated 
with baryon-meson-meson states.
These corrections are, in general, small due to
the suppression associated with the multiple meson excitations.
One can exemplify this effect using the nucleon case as an example.
The nucleon meson cloud is dominated by the
$\ket{N \pi}$ with $N \equiv qqq$ component.
Based on previous work~\cite{AxialFF}, we can estimate 
the probability of the $\ket{N \pi}$ state as $1- Z_N \approx 0.3$.
The correction associated with the two pion
plus $N=qqq$ state ($\ket{N \pi \pi}$)
contribution is then attenuated by the factor $(1- Z_N)^2  \approx 0.09$.

The decomposition (\ref{eqBaryonS}) is model dependent
since the contributions from the meson cloud cannot be disentangled in experiments.
The model dependence is a consequence of the relation
between the meson cloud contributions and the scale associated
to the bare core~\cite{Bernard98a,Hammer04a,Meissner07a}.
In Sec.~\ref{secMC}, we discuss how the model
dependence can be reduced by the comparison with
lattice QCD simulations with large pion masses~\cite{Spectator-Review,NDelta,DecupletDecays}.

%%%%%%%%%%%{tablePHI}

\begin{table*}[t]
\begin{center}
\begin{tabular}{l c c c}
\hline
\hline
$B$   & $\ket{M_A}$  & &  $\ket{M_S}$  \\
\hline
$p$     &  $\sfrac{1}{\sqrt{2}} (ud -du) u$ & &
        $\sfrac{1}{\sqrt{6}} \left[
        (ud + du) u - 2 uu d \right]$ \\
$n$     & $\sfrac{1}{\sqrt{2}} (ud -du) d$  & &
         $-\sfrac{1}{\sqrt{6}} \left[
         (ud + du) d - 2 ddu \right]$  \\
\hline
$\Lambda^0$ &
$\sfrac{1}{\sqrt{12}}
\left[
s (du-ud) - (dsu-usd) + 2(ud -du)s
\right]$
& &
$\sfrac{1}{2}
\left[ (dsu-usd) - s (ud-du)
\right]$ \\
\hline
$\Sigma^+$  &  $\sfrac{1}{\sqrt{2}} (us -su) u $ & &  
$\sfrac{1}{\sqrt{6}} \left[(us + su) u - 2 uu s \right]$
\\
$\Sigma^0$ &
$\sfrac{1}{2}
\left[ (dsu+usd) -s (ud+du)
\right]$
& &
$\sfrac{1}{\sqrt{12}}
\left[
s (ud + du ) +(dsu+usd) -2(ud+du)s
\right]$ \\
$\Sigma^-$ & $\sfrac{1}{\sqrt{2}} (ds -sd) d$ & &
$\sfrac{1}{\sqrt{6}}\left[ (sd + ds) d - 2 dd s \right]$ 
              \\
\hline
$\Xi^0$ & 
$\sfrac{1}{\sqrt{2}} (us -su) s$ 
& &
$-\sfrac{1}{\sqrt{6}} \left[(us + su) s - 2 ss u\right]$   \\
$\Xi^-$ & $\sfrac{1}{\sqrt{2}} (ds -sd) s$
& &
$-\sfrac{1}{\sqrt{6}} \left[(ds + sd) s - 2 ss d\right]$  \\
\hline
\hline
\end{tabular}
\end{center}
\caption{ \footnotesize{Representations 
of the flavor wave functions of the octet baryons.}}
\label{tabPHI}
\end{table*}

\section{Model for the vacuum \label{secVacuum}}

We discuss here the framework of the covariant spectator quark model 
used in the present work for the study of the
electroweak structure of baryons in vacuum.

The covariant spectator quark model is a 
constituent quark model derived from 
the covariant spectator theory~\cite{Nucleon,Spectator-Review,Gross}.
Within the framework, the baryons are described 
as three-constituent quark systems, where a quark 
is free to interact with electroweak fields in the impulse approximation.
The electromagnetic interaction with the quarks
is described in terms of quark electromagnetic and axial 
form factors, that simulate the structure associated with
the gluon and quark-antiquark dressing of the quarks.
The constituent quark form factors are parametrized 
using a vector meson dominance picture
calibrated in the study of the electromagnetic structure 
of the nucleon and baryon decuplet~\cite{Nucleon,Omega}.

In the electroweak interaction with the baryon systems
the probes act on the individual quarks in
the relativistic impulse approximation.
In the covariant spectator theory formalism~\cite{Gross}
one can then integrate over the quark-pair degrees of freedom
since they are spectators in the interaction,
and reduce the system to a quark-diquark state,
where the diquark can be represented 
as an on-mass-shell particle with effective mass
$m_D$~\cite{Nucleon,Nucleon2}. 
The baryon wave functions are derived from 
spin-flavor-radial SU(6)$\otimes$O(3) symmetry associated
with the quark-diquark configurations.
The radial wave functions are determined phenomenologically
by experimental data, or lattice QCD results for some 
ground state systems~\cite{Nucleon,Omega,Lattice,Octet,Octet2,OctetDecuplet1,Roper}.
In the model, the SU(3) quark flavor symmetry-breaking effect
is reflected at the level of the baryon radial wave functions
considering different range parameters for the systems with zero,
one and two strange quarks, as discussed next.

We can now summarize the results from Ref.~\cite{AxialFF} 
for the octet baryon axial form factors $G_A$ and $G_P$ in the vacuum.

We consider the decomposition 
of the axial-vector $G_A$ and induced pseudoscalar $G_P$ 
form factors, according to
\ba
& &
G_A = G_A^{\rm B} + G_A^{\rm MC}, 
\label{eqGA1}\\
& &
G_P = G_P^{\rm B}  + G_P^{\rm pole} +  G_P^{\rm MC}. 
\label{eqGP1}
\ea 
The pseudoscalar meson pole contribution to $G_P$ can be defined as
\ba
G_P^{\rm Pole} = 
\frac{4M_{BB'}^2}{\mu^2 + Q^2} G_A^{\rm B},
\label{eqGP-pole}
\ea
where $\mu = m_\pi$ (pion mass) for $|\Delta I|=1$ transitions 
and   $\mu = m_K$ (kaon mass) for $|\Delta S|=1$ transitions,
as discussed in Ref.~\cite{AxialFF}. 

The explicit meaning of these bare (B) and meson cloud (MC) 
contributions are explained in the following sections.

\subsection{Bare contribution}

Our starting point is the covariant spectator quark model 
for the nucleon and octet baryons developed for the
study of the electromagnetic structure
of those systems~\cite{Nucleon,Octet}.
The main difference between the two works is that in the study 
of the weak axial structure we consider, in addition to the
dominant $S$-state, a $P$-state contribution to
the octet baryon quark-diquark wave functions~\cite{AxialFF,Nucleon2}:
\ba
\Psi_B(P,k) =  n_S \Psi_S (P,k) + n_P \Psi_P (P,k),
\label{eqPsiB}
\ea
where $P$ is the momentum of the baryon (mass $M_B$),
$n_P$ is the $P$-state admixture coefficient
and $n_P = \sqrt{1 - n_S^2}$.

The $S$- and $P$-state wave functions are defined as~\cite{AxialFF}
\ba
& &
\hspace{-.8cm}
\Psi_S(P,k) =
\frac{1}{\sqrt{2}}
\left[ \phi_S^0 \left| M_A \right> 
+ \phi_S^1 \left| M_S \right> 
 \right] \psi_S(P,k), \\
& &
\hspace{-.8cm}
\Psi_P(P,k) = 
{\not \! \tilde k}
\frac{1}{\sqrt{2}}
\left[ \phi_S^0 \left| M_A \right> 
+ \phi_S^1 \left| M_S \right> 
 \right] \psi_P(P,k),
\ea
where $\phi_S^{0,1}$ are the spin wave functions 
labeled by the diquark spin ($S=0,1$), 
$\left| M_{S,A} \right>$ are the flavor states,
symmetric ($S$) or antisymmetric ($A$) 
in the exchange of the flavors between the
quarks 1 and 2, $\psi_S$ and $\psi_P$ 
are the radial wave functions and 
$\tilde k= k - \frac{P \cdot k}{M_B^2} P$.
The flavor states $\left| M_{S,A} \right>$
for the octet baryons are presented in Table~\ref{tabPHI}.

The operator ${\not \! \tilde k}$ is included 
to generate a $P$-state wave function 
based on the structure of a 
$S$-state wave function~\cite{AxialFF,Nucleon2}.
In our calculations, we consider the correlation 
between $S$- and $P$-state radial wave functions
given by $\psi_P(P,k) = N_P \psi_S(P,k)/\sqrt{- \tilde k^2}$,
where $N_P$ is a normalization constant.
In this case the $P$-state radial wave function
is described by the same momentum range parameters 
as the $S$-state.

As discussed in Ref.~\cite{AxialFF} the inclusion of the 
$P$-state is very important for the description of the 
lattice QCD data for the nucleon and the consequent 
extension of the calibration from lattice to the physical regime.

For the study of the weak axial structure of the baryon
we consider the quark axial current operator
\ba
j_{Aq}^\mu = 
\left(  g_A^q (Q^2) \gamma^\mu +
g_P^q(Q^2) \frac{q^\mu}{2 M_N}
\right) \gamma_5 \frac{\lambda_a}{2},
\label{eqJAq}
\ea
where $\lambda_a$  ($a=1,...,8$) are the Gell-Mann  matrices
and $g_A^q$ and $g_P^q$ are the quark 
axial-vector and induced pseudoscalar form factors, respectively.
In Eq.~(\ref{eqJAq}) the flavor operators $\lambda_a$
act on the quark flavor states ($\left|M_{A,S} \right>$) and the Lorentz
operators on the baryon spin states.
The explicit form of the matrices $\lambda_a$
are presented in Appendix~\ref{appGM-matrices}. 
Notice in Eq.~(\ref{eqJAq}) the inclusion of the nucleon
mass in the definition of the quark current.
This property, used already in the definition
of the electromagnetic current~\cite{Nucleon,Omega}
is fundamental for the extension of the model
to the lattice QCD regime.

As discussed in Ref.~\cite{AxialFF}, we  consider
the following parametrization for the
quark form factors
\ba
& &
\hspace{-1cm}
g_A^q (Q^2)= \lambda + (1-\lambda)\frac{m_\rho^2}{m_\rho^2+ Q^2}
+ c_-  \frac{ Q^2 M_h^2}{(M_h^2+ Q^2)^2}, \label{gqA}\\
& &
\hspace{-1cm}
g_P^q (Q^2)= \alpha  \frac{m_\rho^2}{m_\rho^2+ Q^2} +
\beta  \frac{M_h^2}{M_h^2+ Q^2},
\label{eqQuarkGP}
\ea
where $m_\rho$ is the $\rho$ meson mass and
$M_h$ represents a mass of an effective heavy meson
that simulates the structure of all the 
high mass resonances. We consider $M_h= 2 M_N$, 
where $M_N$ is the nucleon mass.
The parameters $\lambda$, $c_-$, $\alpha$ and $\beta$
were determined in previous studies~\cite{AxialFF,Octet2}.

In the covariant spectator quark model the  
radial wave functions of the baryon $B$ 
are represented in terms of 
\ba
\chi_{_B}= \frac{(M_B-m_D)^2-(P-k)^2}{M_B  m_D},
\label{eqChiN}
\ea
where $M_B$ is the mass of the baryon.

The radial wave functions of the 
octet baryon $S$-states take the form 
\ba
\psi_B (P,k) = \frac{N_B}{m_D (\beta_1 + \chi_{_B}) (\beta_l + \chi_{_B}) },
\label{eqPsiB-rad}
\ea
where $l=2$ ($N$), $l=3$ ($\Lambda$ and $\Sigma$) or $l=4$ ($\Xi$).
Numerically we use the parametrization from Ref.~\cite{Octet2}.
Notice that $\beta_2 > \beta_3 > \beta_4$ 
reproduce the natural size of the baryons, 
where systems with more strange quarks are more compact.

In the covariant spectator quark model the weak transition form factors
are determined in the relativistic impulse approximation
by the current~\cite{Nucleon,Omega,Nucleon2}
\ba
\hspace{-1.cm}
(J_5^\mu)_{B'B} =
3 \sum_{\Gamma}
\int_k \overline \Psi_{B'}(P_+,k) (j_{Aq}^\mu) \Psi_B(P_-,k), 
\label{eqJ5}
\ea
where the index $a$ in Eq.~(\ref{eqJAq}) is determined 
by the flavors associated with the transition $B \to B'$.
The integration symbol represent the
integration on the diquark on-shell momentum.

In the previous equation we consider the axial coupling
with the single quark (label 3) when the spectator
quarks have labels 1 and 2 [diquark (12)].
The remaining contributions
associated with the diquarks (13) and (23)
are identical due to the symmetry in the change of the quark labels.
The final contribution can then be obtained
considering the contribution of a single diquark state
multiplied by the factor 3~\cite{Nucleon,Nucleon2}.
The final expression for the transition current (\ref{eqJ5})
based on a single quark operator
(i.e.~no exchange or interaction currents) is valid
when we consider a phenomenological quark-diquark
radial wave function %(\ref{eqPsiB-rad})
with effective parameters determined by physical or lattice QCD data.
A more detailed derivation of the relativistic impulse approximation
within the covariant spectator quark model framework
can be found in Ref.~\cite{Nucleon2}.

For the determination of the axial form factors 
we assume that the current (\ref{eqJ5}) can
be written in the form
\ba
& &
(J_5^\mu)_{B'B} =  \nonumber \\
& & \frac{1}{2}
\bar u(P_+)
\left[
\tilde G_A^{\rm B} (Q^2) \gamma^\mu + 
\tilde G_P^{\rm B}(Q^2) \frac{q^\mu}{2M_{BB'}}
\right] \gamma_5 u(P_-), \nonumber \\
\label{eqGAbare}
\ea
which defines the bare contributions $\tilde G_A^{\rm B}$
and $\tilde G_P^{\rm B}$ to the axial form factors.
These functions do not include the effect 
of the normalization of the wave functions,
which are discussed later.

With the form (\ref{eqPsiB-rad}) chosen for the radial
wave functions, including the factor $1/m_D$, the transition form factors
are independent of the diquark mass~\cite{Nucleon,Omega}.
As mentioned already, the SU(3) symmetry breaking due
to the mass of the strange quark is taken into account
in the range parameters of the wave functions determined
by the global fit to the octet baryon
electromagnetic form factors~\cite{Octet2}.

\begin{table*}[t]
\begin{center}
\begin{tabular}{c c | c c c  c c}
\hline
\hline
  &   & $B \to B'$&   $f_{X}^A$ & $f_{X}^S$ & & \hspace{2em}${\cal F}$\\
\hline 
$|\Delta I| = 1 \hspace{1em}$ 
&  ($I_+$) & \hspace{2em}$n \to p$\hspace{2em}                  
      &$1$ &$-\frac{1}{3}$ && \hspace{2em} $\frac{10}{9}$ \\
   &  $(I_\mp)$          &\hspace{2em}$\Sigma^{\pm} \to \Lambda$\hspace{2em}  
&$\pm\frac{1}{\sqrt{6}}$ &$\mp\frac{1}{\sqrt{6}}$
&& \hspace{2em}   $\pm  \frac{2\sqrt{2}}{3\sqrt{3}}$\\
   &   $(I_+)$             &\hspace{2em} $\Sigma^- \to \Sigma^0$\hspace{2em}
&$\frac{1}{\sqrt{2}}$ &$\frac{1}{3\sqrt{2}}$
&& \hspace{2em} $\frac{4\sqrt{2}}{9} $ \\
   &   $(I_+)$             &\hspace{2em} $\Sigma^0 \to \Sigma^+$\hspace{2em}
&$\frac{1}{\sqrt{2}}$ &$\frac{1}{3\sqrt{2}}$
&& \hspace{2em} $\frac{4\sqrt{2}}{9} $ \\
   &  $(I_+)$            &\hspace{2em}$\Xi^- \to \Xi^0$\hspace{2em} 
  &$0$   &$\frac{2}{3}$ && \hspace{2em} $- \frac{2}{9}$\\
\hline
$|\Delta S| = 1\hspace{1em} $ & 
($V_+$)  &\hspace{2em}$\Lambda \to p$\hspace{2em} 
                                & 
  $-\frac{2}{\sqrt{6}}$ & $0$  &&\hspace{2em} $- \sqrt{\frac{2}{3}}$ \\
   & $(V_+)$                      &\hspace{2em}$\Sigma^{-} \to n$\hspace{2em} 
          &$0$ &  $-\frac{2}{3}$ && \hspace{2em} $\frac{2}{9}$ \\
   & $(V_+)$                      &\hspace{1.3em} $\Sigma^0 \to p$\hspace{2em} 
          &$0$ &  $-\frac{\sqrt{2}}{3}$  && \hspace{2em} $\frac{\sqrt{2}}{9}$\\

   & $(V_+)$                 &\hspace{2em}$\Xi^- \to \Lambda$\hspace{2em} 
&$-\frac{1}{\sqrt{6}}$ &  $-\frac{1}{\sqrt{6}}$
&&  \hspace{2em} $- \frac{\sqrt{2}}{3\sqrt{3}}$\\
   &  ($V_+$)                 &\hspace{2em}$\Xi^- \to \Sigma^0$\hspace{2em} 
           &$\frac{1}{\sqrt{2}}$ &$-\frac{1}{3\sqrt{2}}$  && \hspace{2em}  $\frac{5 \sqrt{2}}{9}$\\
    & ($V_+$)                 &\hspace{2em}$\Xi^0 \to \Sigma^+$\hspace{2em} 
&$1$ &$-\frac{1}{3}$  &&\hspace{2em} $\frac{10}{9}$ \\
\hline
%$\Delta I = 0$,
$\Delta I = 0, \hspace{1em}$
& ($I_0$)  &    $N \to N$ \hspace{2em}&  $\tau_3$ & \hspace{0em} $- \frac{1}{3} \tau_3$ &&
\hspace{2em}  $\frac{10}{9} \tau_3$\\
$\Delta S = 0  \hspace{1em}$
& ($I_0$)  & \hspace{1.5em}  $\Sigma \to \Sigma $ \hspace{2em}  &  \hspace{2em} $I_\Sigma$  \hspace{2em} &
\hspace{1em}$\frac{1}{3} I_\Sigma$  && \hspace{2em} $\frac{8}{9} I_\Sigma$ \\
& ($I_0$) &  \hspace{1.5em} $\Xi \to \Xi$ \hspace{2em}  &  \hspace{0em} $0$ \hspace{2em}&
\hspace{0.7em} $\frac{2}{3} \tau_3$  && \hspace{2em} $-\frac{2}{9} \tau_3$\\
\hline
\hline
\end{tabular}
\end{center}
\caption{\footnotesize{Coefficients $f_{I}^{S,A}$ and $f_{V}^{S,A}$ for the octet baryon transitions.
The last three lines are associated with the neutral transitions.
The operators act on the isospin states $p= \Xi^0 = \left( 1 \; 0 \right)^T$
($T$: transposition)
and $n= \Xi^- = \left( 0 \; 1 \right)^T$ for isospin $1/2$ and
$I_\Sigma= \mbox{diag}(1,0,-1)$ act on the isospin 1 states (charge operator).
In the last column we include the factor ${\cal F}$.}}
\label{tabFAS}
\end{table*}

The results obtained for the form factors 
for the wave functions (\ref{eqPsiB}) 
and current (\ref{eqJAq}) are
\ba
\tilde G_A^{\rm B} &=&  
g_A^q {\cal F}
\left\{
\frac{3}{2} n_{S}^{2} B_0 
- 3 n_{SP}  \frac{\tau}{1 + \tau}B_1 \right. \nonumber \\
& &
\left. + \frac{6}{5} n_P^{2} 
\left[ \tau B_2 -(1+ \tau) B_4 \right]
\right\}, \label{eqGAoct} \\
\tilde G_P^{\rm B} &=& 
g_A^q {\cal F}
\left\{
 -  3 n_{SP} \frac{1}{1 + \tau} B_1 \right. \nonumber \\
& &
\left. + \frac{3}{2} n_P^{2} 
\left[ \frac{B_5}{\tau} + 2(B_2-B_4) \right]
\right\} \nonumber \\
& &
+ \frac{M_{BB'}}{M}g_P^q {\cal F}
\left\{
\frac{3}{2} n_{S}^{2} B_0 - 
3 n_{SP} B_1 \right. \nonumber \\
& &
\left. + \frac{3}{2} n_P^{2} 
\left[ \tau B_2 +  B_3   - (2 + \tau)B_4 \right]
\right\},
\label{eqGPoct}
\ea
where $n_{SP}= n_S n_P$, 
$\tau= \frac{Q^2}{4 M_{BB'}^2}$, and  
the functions $B_i$ ($i=0,...,5$) 
are integrals that include 
the radial wave functions $\psi_S$ and $\psi_P$
of the initial and final states.

The factor ${\cal F}$ take the form
\ba
{\cal F}= f_{X}^A - \frac{1}{3} f_{X}^S,
\ea
where the coefficients $f_{X}^{A,S}$ are determined by
the projections of the flavor transition operators ($X=I_\pm$ or 
and $X=V_\pm$) on the flavor states
\ba
& &
f_{X}^A = 
{_{B'}}\!\left< M_A | X | M_A \right> _B, \\
& &
f_{X}^S = 
{_{B'}}\!\left< M_S | X | M_S \right> _B.
\ea 
The flavor transition operators are discussed in Appendix~\ref{appGM-matrices}. 
The coefficients $f_{X}^{A,S}$ and the factor ${\cal F}$ are presented in Table~\ref{tabFAS}.

In the previous equations $\tilde G_A$, $\tilde G_P$,
$g_A^q$, $g_P^q$ and $B_i$ are functions of $Q^2$.
The explicit expressions for $B_i$ ($i=0,...,5$) can be found in Ref.~\cite{AxialFF}.
The normalization of the radial wave functions leads to $B_0(0)=1$.
The effect of the mass ($M_N$ or $M_{BB'}$) on the transition form factors 
appears on the functions $B_i$.
The dependence of the transition form factors on the mass of
the baryons was studied in detail in Ref.~\cite{AxialFF},
demonstrating the relevance of taking into account implicitly
the effects of the SU(3) flavor-symmetry breaking.

The leading-order contributions for $\tilde G_A^{\rm B}$ and $\tilde G_P^{\rm B}$ comes from
the $S$-state contribution.
In the $S$-state limit ($n_P= n_{SP}=0$) the form factors
are proportional to $\frac{3}{2}{\cal F}$ and we recover nonrelativistic results.
For the nucleon $\tilde G_A^{\rm B} \simeq \frac{5}{3} g_A^q B_0$.
Different SU(3) coefficients are obtained for the other transitions.

In Table~\ref{tabFAS}, the transitions $|\Delta I| =1$ correspond to 
$d \to u$ transition, except for the $\Sigma^+ \to \Lambda$ ($u \to d$ transition).
From the table, 
we can conclude that the $\Sigma^+ \to \Lambda$
and $\Sigma^- \to \Lambda$ transition form factors differ by a sign,
and that the $\Sigma^0 \to \Sigma^+$ and $\Sigma^- \to \Sigma^+$
transition form factors are identical.
One can then conclude that in the case $|\Delta I| =1$
there are only four cases of interest.

For the study of the $Q^2$ dependence
of the form factors at large $Q^2$ it is convenient to
determine the asymptotic form of the different components
of the axial form factors.
At large $Q^2$ we expect that the bare form factors follow:
\ba
G_A^{\rm B} \propto \frac{1}{Q^4},
\hspace{1cm}
G_P^{\rm B} \propto \frac{1}{Q^6}.
\label{eq-AssFF}
\ea
These results are the consequence of the asymptotic form from
the functions $B_i$ and that $g_A^q \propto 1$ and $g_P^q \propto 1/Q^2$.
From the expressions in Ref.~\cite{AxialFF}
one can conclude that $B_0,B_1 \propto 1/Q^4$
and that  $B_i \propto 1/Q^6$ for $i=2,3,4,5$.

\subsection{Meson cloud contribution \label{secMC}}

We consider now the effect of the meson cloud.
Following Ref.~\cite{AxialFF}, we consider a phenomenological
description of the meson cloud effects.
The use of a phenomenological description
was motivated by the fact that the magnitude
of the meson cloud contributions is expected to
be smaller than the valence quark contributions~\cite{AxialFF,Octet,Octet2},
and by simplicity.

When the baryon wave function has contributions from
valence quarks and from the meson cloud  
we can consider the following decomposition
\ba
\hspace{-1cm}
G_A (Q^2) 
= \sqrt{Z_{B'}Z_B} 
\left[ \tilde G_A^{\rm B}(Q^2) + \tilde G_A^{\rm MC} (Q^2)  \right],
\label{eqGA-decomp}
\ea 
where $Z_B$ represents the normalization factor determined 
in the study of the baryon $B$ electromagnetic factors~\cite{Octet2}.
The bare contribution is associated with the short-range interactions,
and the meson cloud contribution is associated
with the long-range interactions.

From the comparison with Eqs.~(\ref{eqGA1}) and (\ref{eqGP1})
we can write 
$G_A^{\rm B}=  \sqrt{Z_{B'}Z_B}  \tilde G_A^{\rm B}$
and  $G_A^{\rm MC}= \sqrt{Z_{B'}Z_B}  \tilde G_A^{\rm MC}$.
In the transition $n \to p$, 
we can replace $\sqrt{Z_{B'}Z_B}  \to Z_N$.

As mentioned already, the decomposition from Eq.~(\ref{eqGA-decomp})
into a bare and meson cloud contributions is model dependent~\cite{Ramalho-Pena23,Pascalutsa07a}.
This model dependence is a consequence of the ambiguities on the calibration
of the background and in the identification of
the bare states~\cite{Bernard98a,Hammer04a,Meissner07a,Ronchenn13a,Pascalutsa07a}.
The impact of the model dependence can be reduced by matching the valence
quark contributions of our model with the results from lattice QCD simulations
for large pion masses, as explained next.
Although quenched and unquenched lattice QCD simulations include
some meson cloud effects, those effects are reduced for larger 
pion masses~\cite{Pascalutsa07a,Alexandrou08a}.

The meson cloud contribution to $G_P$ is determined by 
pole contribution associated with $G_A^{\rm MC}$
\ba
G_P^{\rm MC} (Q^2) = 
\sqrt{Z_{B'}Z_B} \frac{4M_{BB'}^2}{\mu^2 + Q^2} \tilde G_A^{\rm MC} (Q^2),
\label{eqGPMC}
\ea
where $\mu = m_\pi$ (pion mass) for $|\Delta I| =1$ transitions 
and  $\mu = m_K$ (kaon mass) for $|\Delta S| =1$ transitions,
as discussed in Ref.~\cite{AxialFF}.

To simplify the terminology, we call the tilde quantities
($\tilde G_A^{\rm B}$, $\tilde G_P^{\rm B}$, $\tilde G_A^{\rm MC}$, 
$\tilde G_P^{\rm pole}$ and $\tilde G_A^{\rm MC}$, 
bare, MC or pole) as relative contributions
to the form factors, and the relations corrected 
by the wave function normalization factors
as the effective contributions to the form factors.
Notice that the final result requires the factor 
  $\sqrt{Z_{B'}Z_B}$, while the bare contributions are 
naturally independent of these factors.
At large $Q^2$ the meson cloud contribution is suppressed 
and $G_A \simeq \sqrt{Z_{B'} Z_B} \tilde  G_A^{\rm B}$.
In the present work, we assume that the dominant contribution
is the pion cloud and use the relations from Ref.~\cite{Octet2}
to determine $Z_B$.

The normalization constant $Z_B$ is determined by the 
study of the octet baryon electromagnetic
form factors~\cite{Octet,Octet2}, and can be represented as
\ba
Z_B = \frac{1}{1 + 3 a_B b_1},
\label{eqZB1}
\ea
where $b_1= 0.121$ and $a_N= 1$, $a_\Lambda = 0.48$, 
$a_\Sigma =0.59$ and $a_\Xi= 0.04$.
The values for $a_B$ are the result of 
the SU(3) structure of the pion-baryon interaction.
The coefficient for $\Sigma$ include the intermediate 
couplings with $\Sigma$ and $\Lambda$ in an effective way.

We consider here the calibration derived in the study of the axial-vector form factors.
More specifically $Z_N$ was determined by the comparison 
of $Z_N \tilde G_A^{\rm B}$ with the experimental parametrization of 
$G_A$ for $Q^2 > 1$ GeV$^2$ (large $Q^2$ region)~\cite{AxialFF}.

\begin{figure}[t]
\vspace{1.2cm}
  \includegraphics[width=2.in]{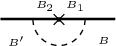}
\caption{\footnotesize{Diagram associated with the
meson cloud interaction in leading order.
\label{figMesonCloud}}}
\end{figure}

In the representation (\ref{eqZB1}) 
the explicit dependence of the pion cloud contributions 
on the coupling constants $g_{\pi B B'}$ is transferred to the factor $b_1$
since the coefficients $a_B$ are normalized by the result for the nucleon.
Since in the electromagnetic interaction with the pion cloud
in leading order includes two pion-baryon-baryon vertices
(processes similar to Fig.~\ref{figMesonCloud}),
associated with the coupling constants $g_{\pi B B'}$,
and that in an SU(3) meson-baryon interaction
the coupling constants are proportional to $g_{\pi NN}$,
we can write
\ba
b_1 \propto g_{\pi NN}^2.
\ea

We can now discuss the function $\tilde G_A^{\rm MC}$
which determines the meson cloud contribution to the form factors $G_A$.
For $\tilde G_A^{\rm MC}$, we consider an effective parametrization 
based on an SU(3) meson-baryon coupling,
with two independent functions $F'$ and $D'$ of $Q^2$~\cite{Gaillard84,AxialFF}.
We use primes to emphasize the difference to the standard
SU(3) meson-baryon model, where the functions parametrize
the combination of valence and meson cloud contributions.
The value of the functions for $Q^2=0$ 
can be determined by the  
results for $G_A(0)$ for different known transitions 
($n \to p$, $\Lambda \to p$, $\Sigma^- \to n$, 
$\Xi^- \to \Lambda$ and $\Xi^0 \to \Sigma^+$).
The explicit values are 
\ba
\hspace{-1.cm}
F_0 \equiv F' (0)= 0.1775, \hspace{.4cm} D_0 \equiv D'(0) = 0.4284. 
\label{eqFD}
\ea
For convenience we define also $H_0= F_0 + D_0$.

In the extension from the nucleon to the octet baryon cases
we replace $F'$ and $D'$ by functions of $Q^2$,
according to the $Q^2$ dependence observed 
for the meson cloud contribution to the $n \to p$ transition, 
as described next.

We emphasize that we consider here {\it a modified version}, 
where, only the meson cloud contributions are described by 
a meson-baryon SU(3) symmetry model.
The bare contributions are determined by our approximated 
SU(6) quark model, where the breaking is included
at the level of the radial wave functions 
with parameters depending on the number of strange quarks,
and by the values of the baryon masses.

\begin{table*}[t]
\begin{center}
\begin{tabular}{c | c c c }
\hline
\hline
  &Process & $\tilde G_A^{\rm MC}$ &  $\eta_{BB'} $\\
\hline
$|\Delta I| = 1\hspace{2em}$ &\hspace{2em}$n \to p$\hspace{2em} &$F'+D'$  & $\frac{F_0 + D_0}{H_0}$\\
                           &\hspace{2em}$\Sigma^{\pm} \to \Lambda$\hspace{2em} &$ \pm \sqrt{\frac{2}{3}}  D'$ 
&   $ \pm \sqrt{\frac{2}{3}}  \frac{D_0}{H_0} $ \\
                           & \hspace{2.5em}$\Sigma^{-} \to \Sigma^0$ \hspace{2em} & $\sqrt{2} F'$
&
 $\sqrt{2} \frac{F_0}{H_0}$  \\
                           & \hspace{2.5em}$\Sigma^{0} \to \Sigma^+$ \hspace{2em} & $\sqrt{2} F'$
&
 $\sqrt{2} \frac{F_0}{H_0}$  \\
                           &\hspace{2em}$\Xi^- \to \Xi^0$\hspace{2em} &$F'-D'$  &  
$ \frac{F_0 -D_0}{H_0}$\\
\hline
$|\Delta S| = 1\hspace{2em}$ &\hspace{2em}$\Lambda \to p$\hspace{2em} &
$-\sqrt{\frac{3}{2}}  \left(F' + \frac{1}{3}D'\right)$ 
&  $-\sqrt{\frac{3}{2}}  \frac{F_0 + D_0/3}{H_0}$  \\
% $-\sqrt{\frac{3}{2}}  \left(F_0 + \frac{1}{3}D_0\right)$  \\
                           &\hspace{2.em}$\Sigma^{-} \to n$\hspace{2em} &$-F'+D'$  & 
$- \frac{F_0 -D_0}{H_0}$ \\
                     &\hspace{1.6em}$\Sigma^{0} \to p$\hspace{2em} &
                       $-\frac{1}{ \sqrt{2}} (F'-D')$  & 
  $-\frac{1}{ \sqrt{2}} \frac{F_0 -D_0}{H_0}$   \\
                         &\hspace{2em}$\Xi^- \to \Lambda$\hspace{2em} &
$- \sqrt{\frac{3}{2}} \left(F'-\frac{1}{3}D'\right)$  &
$- \sqrt{\frac{3}{2}}  \frac{F_0 - D_0/3}{H_0}$ \\
                           &\hspace{2.3em}$\Xi^- \to \Sigma^0$\hspace{2em} &
$\frac{1}{\sqrt{2}} (F' + D')$ & 
$\frac{1}{\sqrt{2}} \frac{F_0 + D_0}{H_0}$  \\
                           &\hspace{2.2em}$\Xi^0 \to \Sigma^+$\hspace{2em} &$F'+D'$ &  $\frac{F_0 + D_0}{H_0}$\\
\hline
$\,\Delta I=0, \hspace{2em}$  & $N \to N$    &      $(F'+D')\tau_3$ & $ \frac{F_0 + D_0}{H_0} \tau_3$ \\
$\Delta S=0 \hspace{2em}$  & $\Sigma \to \Sigma$ & $2F' I_\Sigma$ &
$2 \frac{F_0}{H_0} I_\Sigma$ \\
              &  $\Xi \to \Xi$ & $ (F' - D')\tau_3$ & $ \frac{F_0 - D_0}{H_0} \tau_3$  \\
\hline
\hline
\end{tabular}
\end{center}
\caption{\footnotesize{Octet baryon axial-vector form factors $G_A(Q^2)$, 
expressed in terms of $F^\prime$ and $D^\prime$
of SU(3) scheme~\cite{Gaillard84,Lin09,Erkol}.
$F_0$ and $D_0$ are defined by Eq.~(\ref{eqFD})
and $H_0= F_0 + D_0$.
The last rows refer to the neutral currents.
The contributions to $\Lambda$ and $\Sigma^0$ are zero~\cite{AxialFF}.}
}
\label{tabGA}
\end{table*}

The meson cloud contribution for the nucleon axial-vector form 
factor is determined by the fit to the nucleon data
based on the results for $G_A$ estimated by
the lattice QCD data
\ba
\tilde G_{A,N}^{\rm MC} (Q^2) =
\frac{G_A^{{\rm MC}0}}{\left( 
1 + \frac{Q^2}{\Lambda^2}\right)^4},
\label{eqGAmc}
\ea
where $G_A^{{\rm MC}0}= 0.6059$
and $\Lambda= 1.05$ GeV.
In the notation of Table~\ref{tabGA},  $G_A^{{\rm MC}0} =  H_0$.
In Ref.~\cite{AxialFF}, we demonstrated the advantage
of using a SU(3) meson-baryon parametrization to describe 
the meson cloud component of the model.

%%%%    Figure 2

The generalization of the meson cloud to 
an arbitrary $B \to B'$ transition  
follows the form
\ba
\tilde G_A^{\rm MC}(Q^2)= \eta_{BB'}
\;  \tilde G_{A,N}^{\rm MC}(Q^2),
\label{eqGMCoctet}
\ea 
where $\tilde G_{A,N}^{\rm MC}$ represents 
the parametrization of the nucleon's meson cloud contribution
given by Eq.~(\ref{eqGAmc})
and $\eta_{BB'}$ is the coefficient 
associated with the SU(3) symmetry
displayed in the last column in Table~\ref{tabGA}. 
Notice that the coefficient $\eta_{BB'}$ is a constant 
defined by the values of $F_0$ and $D_0$ 
and the $Q^2$ dependence comes exclusively
from the function $\tilde G_{A,N}^{\rm MC}$.

The effective meson cloud contribution to the 
axial-vector form factors is obtained when we take into account
the normalization factors
\ba
G_A^{\rm MC}(Q^2)= 
\sqrt{Z_{B'} Z_{B}} \; 
\tilde G_A^{\rm MC}(Q^2).
\label{eqGMCoctet2}
\ea

In Table~\ref{tabGA} we represent some factors $\eta_{BB'}$
by using the factor $(F_0 + D_0)/H_0$ when we could have used unity.
This representation is intentional in order to
prepare the generalization for a case
where the meson-baryon coupling may change,
as discussed in the next section
(extension to the nuclear medium).
In that case we consider that the functions $\tilde G_A^{\rm MC}$
change with the (squared) coupling constants.
This can be done by assuming that $F_0$ and $D_0$ include
the dependence on the coupling constants,
while $H_0$ is constant determined numerically
and independent of the coupling constants.

Like the meson cloud contributions, the bare contributions 
are broken by the normalization associated 
with the baryon wave functions (factors $\sqrt{Z_B}$).
In these conditions the meson cloud contributions 
for $G_A$ for the transitions $n \to p$ 
and $\Xi^0 \to \Sigma^+$, determined by $F^\prime + D^\prime$,
are not the same since the suppression associated 
with the factors  $\sqrt{Z_B}$ is stronger for the nucleon
(smaller $\sqrt{Z_B}$).

\begin{figure*}[t]
  \vspace{.6cm}
\begin{center}
  \mbox{
\includegraphics[width=3.2in]{GA-total2} \hspace{.5cm}
\includegraphics[width=3.2in]{GP-np-test}}
\end{center}
\caption{\footnotesize{
Model results for the nucleon form factors $G_A$ and $G_P$ according
to the calibration discussed in the text.
{\bf At the left:}  we compare the $G_A$ total and bare results
with the parametrization of the data (red band) and
the estimate of the bare contribution (blue dashed line)
extrapolated from lattice~\cite{AxialFF,Alexandrou11a}.
{\bf At the right:} we present the results for $G_P$, decomposed into
bare contribution, (bare + pole) and (bare + pole + meson cloud).
The $G_P$ data are from Refs.~\cite{Bernard02,Choi93}.
\label{figGAGP-Nucleon}}}
\end{figure*}

In the present work we consider a modification of 
the parametrization from Ref.~\cite{AxialFF},
in order to correct small numerical errors 
in the calculation of $Z_\Xi$ and $Z_\Sigma$.
We use this opportunity to calibrate
the parametrization for $F_0$ and $D_0$ since
the previous estimates of the $\Xi$ decays 
 were slightly overestimated.
The new parameters are displayed in Eq.~(\ref{eqFD}).
Compared with Ref.~\cite{AxialFF} 
the corrections are very small (less than 0.5\%), 
but we consider the new values for consistency.
The main difference between the calculations 
is due to the factor $Z_\Xi$, underestimated about 10\%.
The numerical results for the axial-vector form factors
in vacuum are presented and discussed in Sec.~\ref{sec-Vacuum}.

\subsection{Combination of bare and meson cloud \label{secBpMC}}

Our model for the octet baryon axial form factors is then
based on Eqs.~(\ref{eqGA1}), (\ref{eqGP1}), (\ref{eqGP-pole}),
(\ref{eqGAoct}), (\ref{eqGPoct}),  (\ref{eqGPMC}) and (\ref{eqGMCoctet})
with the proper normalizations.

The formalism described in the present section
for bare and meson cloud contribution implies
that there is a proportionality
between the transition form factors for
$\Sigma^- \to n$ and $\Sigma^0 \to p$ (factor $1/\sqrt{2}$) and
also between $\Xi^- \to \Sigma^0$ and $\Xi^0 \to \Sigma^+$
(also a factor $1/\sqrt{2}$).
The result can be inferred from the comparison
between the coefficients for the bare contribution (Table \ref{tabFAS})
and meson cloud contributions (Table \ref{tabGA}).
This result also implies that when we compare ratios
between form factors we obtain the same result
for $\Sigma^- \to n$ and $\Sigma^0 \to p$,
as well as for $\Xi^- \to \Sigma^0$ and $\Xi^0 \to \Sigma^+$.

We can now discuss the asymptotic form of the
different components of the axial form factors
(\ref{eqGA1}) and (\ref{eqGP1}).
Based on Eqs.~(\ref{eq-AssFF})  and (\ref{eqGMCoctet}), one has
\ba
G_A^{\rm B} \propto \frac{1}{Q^4}, \hspace{.6cm} G_A^{\rm MC} \propto \frac{1}{Q^8}.
\label{eqGA-ass2}
\ea
The corresponding expressions for $G_P$,
based on Eqs.~(\ref{eqGP-pole}), (\ref{eq-AssFF})  and~(\ref{eqGMCoctet2}) are
\ba
G_P^{\rm B} \propto \frac{1}{Q^6}, \hspace{.5cm} G_P^{\rm pole} \propto \frac{1}{Q^{6}},
 \hspace{.5cm}
 G_P^{\rm MC} \propto \frac{1}{Q^{10}}.
\label{eqGP-ass2}
\ea

The results of the calibration
of the model for the nucleon system are
presented in Fig.~\ref{figGAGP-Nucleon}.
In the following, we present the results
from the extension for the octet  baryons in the vacuum,
and the extension of those results to the nuclear medium.

For the extension of the formalism to the nuclear medium
it is convenient to relate the effective coefficients
$F^\prime$ and $D^\prime$ used in Eqs.~(\ref{eqGMCoctet}) and (\ref{eqGMCoctet2})
to the explicit meson cloud mechanisms 
and the meson-baryon couplings.
From the analysis of the meson cloud diagrams
depicted in Fig.~\ref{figMesonCloud}, one
can conclude that the meson cloud contribution 
is proportional to $g_{MBB_1} g_{M B_1 B'}$ where 
$M$ is the dressing meson ($\pi$, $K$, ...)
and $B_1$ is an intermediate baryon state (octet or decuplet).
Since, in the effective SU(3) meson-baryon interactions
we can write all $g_{MBB'}$ in terms of a coefficient times $g_{\pi NN}$,
we conclude that
\ba
\tilde G_A^{\rm MC} \propto g_{\pi NN}^2.
\ea

The results for the other octet baryon members
are presented in Sec.~\ref{secResults},
where we will discuss also the results for the nuclear medium.

\section{Extension of the model to symmetric nuclear medium \label{secMedium}}

We discuss now the extension of the model for the vacuum described 
in the previous section to symmetric nuclear matter.
The extension is based on the modification of the properties of 
the hadrons and the meson-baryon 
coupling constants to the nuclear matter.
These properties can be estimated by the QMC model 
applied to the symmetric nuclear matter~\cite{QMCReview,Tsushima22a}.

The {\it standard} QMC model which we use is based on the MIT bag model,
to be able to describe the symmetric nuclear matter reasonably well
by using the $\sigma$ and $\omega$ mean fields self-consistently.
The self-consistent exchange of the
Lorentz-scalar-isoscalar $\sigma$ and Lorentz-vector-isoscalar
$\omega$ mean fields among the light quarks in nucleons
provides a good description of the symmetric nuclear matter,
with the incompressibility value of about 300 MeV
(within the extracted empirical range),
without introducing complicated nonlinear higher-order meson interactions.
Based on this rather simple treatment with very less number of parameters,
the QMC model has been applied successfully for many nuclear phenomena
(see e.g., Refs.~\cite{QMCReview,Tsushima22a}).
The in-medium effective masses of hadrons are controlled by
the values of the self-consistently obtained $\sigma$ mean field.
The coupling constants in medium
are estimated using the Goldberger–Treimann relation
and the medium values of the baryon masses, baryon axial-couplings
and pion decay constant $f_\pi$ in medium.

In the extension of our model to nuclear matter, 
we consider the values of the hadron masses (baryons and mesons)
$m_h$ from the vacuum to the medium $m_h^\ast$, the vector potentials,
as well as the modifications of the coupling constants $g_{MBB'}$ in vacuum 
to  $g_{MBB'}^\ast$ in medium.
The consequence of this extension is that the bare contributions 
to the axial form factors are modified, since  
we replace $m_\rho$, $m_\phi$ and $M_h= 2 M_N$, respectively, by
$m_\rho^\ast$, $m_\phi^\ast$ and $M_h^\ast= 2 M_N^\ast $ in the quark axial
current and the radial wave functions of the baryon $B$ are modified
by replacing the baryon mass $M_B$ by the effective baryon mass in medium $M_B^\ast$.
As for the meson cloud contributions, they change
as a consequence of the modification of 
the coupling $g_{MBB'}$ to  $g_{MBB'}^\ast$.

\begin{table}[t]
\begin{center}
\begin{tabular}{l c c c c c c}
\hline 
\hline
$\rho/\rho_0$        &  $M_N^\ast$    & $m_\rho^\ast$ &  $g_{\pi NN}^\ast/g_{\pi NN}$ & $g_A^{N
\ast}/g_A^N$ &  $G_A^N(0)$ &  $n_P$ \\
\hline  
  0         &   0.9390   &  0.7545 &   1.000    & 1.0000   &  1.125     & $-0.5067$     \\
  0.5       &   0.8313   &  0.7061 &   0.921    & 0.9404   &  1.058     & $-0.5371$     \\
  1.0       &   0.7545   &  0.6537 &   0.899    & 0.8920   &  1.003     & $-0.5610$     \\
\hline 
\hline
\end{tabular}
\end{center}
\caption{\footnotesize{Model parameters used in the extension
    to the nuclear medium (nucleon case). 
The last columns include the ratio $g_A^{N\ast}/g_A^N$ used to fix $n_P$ for $\rho > 0$.}}
\label{tab-Medium}
\end{table}

In addition to the bare contributions ($\tilde G_A^{\rm B}$ replaced by $\tilde G_A^{{\rm B}*}$),
the effect of the nuclear medium acts in two levels: 
on the meson cloud contribution
($\tilde G_A^{\rm MC}$ replaced by $\tilde G_A^{{\rm MC}*}$) 
and on the wave function normalization factors $Z_B$ and $Z_{B'}$ 
(global modification of $\sqrt{Z_B Z_{B'}}$ to  $\sqrt{Z_B^\ast Z_{B'}^\ast}$).

We discuss the modifications on the function $\tilde G_A^{{\rm MC}}$ first.
From the discussion from Eq.~(\ref{eqGMCoctet}) relative to 
the meson cloud contribution to a generic $B \to B'$ transition,
we conclude that $\tilde G_A^{\rm MC}$ include linear combinations of $F^\prime$ and $D^\prime$,
which are proportional to the product of two
coupling constants $g_{\pi B B'}^2$, since these include diagrams
with two pion-baryon couplings (see Fig.~\ref{figMesonCloud}).
The combination of coupling constants can be 
expressed in terms of $g_{\pi NN}^2$, as discussed already.
The conclusion is then that  $\tilde G_A^{{\rm MC}} \propto g_{\pi NN}^2$ 
and we can write
\ba
\tilde G_A^{{\rm MC} \ast} = 
\left(\frac{g_{\pi NN}^\ast}{g_{\pi NN}} \right)^2 \tilde G_A^{{\rm MC}}.
\ea

In the extension to the nuclear medium we assume,
for simplicity, that medium effects associated
with the meson cloud contributions are taken into
account using the modified coupling constants and
normalization factors.
Thus, in medium $\tilde G_{A,N}^{\rm MC}$ is determined by 
Eq.~(\ref{eqGAmc}) where the cutoff $\Lambda$ value is determined by
nucleon data in vacuum.

We discuss now the impact of the nuclear medium on the wave function
normalization factors, due to the modification of the meson-baryon couplings.
The discussion is related to the discussion of the factor 
$b_1$ in Eq.~(\ref{eqZB1}).
Since as already discussed, the effect of the coupling constants
affects only $b_1$ and the effect is quadratic, we can write
\ba
b_1^\ast  = \left( \frac{g_{\pi NN}^\ast}{g_{\pi NN}} \right)^2 b_1.
\ea 
The normalization factors are then modified by
\ba
Z_B^\ast = \frac{1}{1 + 3 a_B b_1^\ast},
\label{eqZB1st}
\ea
where the variables $a_B$ keep the values in vacuum.

All the discussions about the medium effects
of $\tilde G_A^{{\rm B}\ast}$ and $\tilde G_A^{{\rm MC}\ast}$,
can be extended to $\tilde G_P^{{\rm B}\ast}$, $\tilde G_P^{{\rm pole}\ast}$
and $\tilde G_P^{{\rm MC}\ast}$.
We just need to keep in mind that,
since pole and meson cloud contributions for $G_P$
(\ref{eqGP-pole}) and (\ref{eqGPMC}) depend on the meson mass ($m_\pi$ or $m_K$),
those masses are replaced by the meson masses in medium $m_\pi^\ast$ or $m_K^\ast$.
In the case of the pion, however,
based on Refs.~\cite{Hayano,Kienle,Meissner,Vogl},
we consider no pion mass modification in nuclear matter ($m_\pi^\ast = m_\pi)$.

The axial-vector form factors of baryon octet in vacuum were
already calculated in Ref.~\cite{AxialFF}.
We consider now the extension to the nuclear medium.
An important property observed in the nuclear medium 
is the suppression of the axial-vector form factors due to
the enhancement of the $P$-state component,
in general associated with the lower component of the quark Dirac spinors.
In the present case the simplest way to include the $P$-state effect 
is to modify the value of $n_P$.
In the absence of a solid, established reference, we adjust the value of $n_P$
in order to reproduce the suppression estimated
by the bag model from Ref.~\cite{Lu01a}.
Since our bare estimate for $G_A^{\rm B}(0)$ for the nucleon
is similar to the Bag Model $g_A^{N}$, we use the
in-medium to free space bare axial-vector coupling
ratio $g_A^{N\ast}/g_A^N$ to fix the amount of $P$-state admixture (coefficient $n_P$).
Notice that the determination of $n_P$ by the values of  $g_A^{N\ast}/g_A^N$
is not an additional constraint of the model.
It is a necessary condition associated with the consistency
with the QMC model.
The QMC model uses the input of the bag model for $g_A^{N\ast}$~\cite{Lu01a}
and this result is used in the calculation
of coupling constants $g^\ast_{M B_1 B_2}$~\cite{Octet2}.
When we include the effect of the meson cloud for the nucleon
on the functions $G_A^\ast$
we modify the values of $G_A(0)$ and $G_A^\ast(0)$, and,
consequently, the ratio 
$G_A^\ast(0)/G_A(0)$ deviates from the $g_A^{N\ast}/g_A^N$.

The reduction of the axial coupling $g_A^N$ in the nuclear medium is also observed in calculations based on the soliton model~\cite{Nyman87a,Meissner89a,Rakhimov98a}.

\begin{table*}[t]
\begin{center}
\begin{tabular}{l| c c c c | c|c c |c c }
\hline 
\hline
$\rho/\rho_0$   &  $M_N^\ast$  & $M_\Lambda^\ast$ & $M_\Sigma^\ast$ &
$M_\Xi^\ast$ & $m_K^\ast$ &$m_\rho^\ast$ &  $m_\phi^\ast$  &$g_{\pi NN}^\ast/g_{\pi NN}$    &  $n_P$ \\
\hline  
  0         &   0.9390   &  1.1160   & 1.1920 & 1.3180 & 0.4937 & 0.7545 &  1.0195 & 1.000  & $-0.507$     \\
  0.5       &   0.8313   &  1.0439   & 1.1214 & 1.2822 & 0.4573 & 0.7061 &  1.0191 & 0.921  & $-0.536$     \\
  1.0       &   0.7545   &  0.9927   & 1.0704 & 1.2567 & 0.4305 & 0.6537 &  1.0189 & 0.899  & $-0.560$     \\
\hline 
\hline
\end{tabular}
\end{center}
\caption{\footnotesize{Model parameters used in the extension to the nuclear medium
(baryon octet case).
The discussion about the values of $n_P$ are in the previous sections.
The kaon mass is used in the calculation of the $G_P$ pole term
in the $|\Delta S| =1$ transitions.
For  $m_\pi^* \simeq m_\pi$ we use $m_\pi= 0.138$
GeV~\cite{Hayano,Kienle,Meissner,Vogl}.}}
\label{tab-Medium2}
\end{table*}

The determination of the value of $n_P$ is based on the
relation~\cite{AxialFF}
\ba
\tilde G_A^{\rm B}(0) = \frac{15-19 n_P^2}{9} g_A^q (0),
\label{eqGAB0}
\ea 
where $g_A^q(0)=1$.

%%%%

In Table~\ref{tab-Medium}, we present the parameters
used in the medium calculations to the axial-vector and 
induced pseudoscalar form factor for the nucleon 
(densities $0.5 \rho_0$ and $\rho_0$).
The coupling constant $g_{\pi N N}^*$ in medium is
calculated based on the Goldberger–Treimann relation, 
the values of $f_\pi^*$, the axial-vector coupling $g_A^{N*}$ and
the nucleon masses $M_N^\ast$~\cite{QMCReview,Octet2,GTrelation,fpiMedium}.
The parameters necessary for the calculations of
the  other transitions between the octet baryons
are presented in the next section.
The values of $f_\pi^*$ in medium are based on Ref.~\cite{fpiMedium},
and are limited to values of $\rho$  below 2.5 $\rho_0$.
The value of $n_P$ is determined by the value of $g_A^{N\ast}$ 
estimated by the ratio included in the table.

%%%%%%%%%    TABLE VI

%%%%  TABLE   VI

%\clearpage

%\newpage

\section{Numerical results \label{secResults}}

In this section we present our numerical results
for the octet baryon axial form factors $G_A$ and $G_P$
in vacuum and in medium.
In the calculations we use the parameters from Table~\ref{tab-Medium2}.

As discussed already, the axial transitions between
octet baryon members correspond to six $|\Delta I|=1$ transitions
(four independent) and six  $|\Delta S| =1$ transitions.
To simplify the analysis and the discussion
of the results, we consider here only typical cases,
for the $|\Delta I| =1$ and $|\Delta S| =1$ channels, 
associated with different octet baryon masses.
The comparison of all channels is presented in Appendixes~\ref{app-GA}
and \ref{app-GP}.

We start with the discussion
of the results for the vacuum and some remarks about
the calibration of our model in vacuum (Sec.~\ref{sec-Vacuum}).
Next we discuss the results in the nuclear medium (Sec.~\ref{sec-Medium1}).
We display the vacuum and in-medium results side by side
in order to better observe the modifications due to the nuclear medium
(Figs.~\ref{figGA-p1}, \ref{figGA-p21}, \ref{figGP-p1} and \ref{figGP-p21}).

%\vspace{1cm}

%\newpage

%%%%%%%%%%%%%%%%%%%%%%%%%%%%%%%%%%%%%%%%%%%%%%%%%%%%%%%%%%%%%%%%%%%%
%
%    FORM FACTORS RATIOS    GA*(Q2)/GA*(0)
%
%%%%%%%%%%%%%%%%%%%%%%%%%%%%%%%%%%%%%%%%%%%%%%%%%%%%%%%%%%%%%%%%%%%%

\begin{figure*}[t]   %%%%    Figure 3 (11)
\vspace{.5cm}
\begin{center}
\mbox{
\includegraphics[width=2.9in]{GA-np} \hspace{1.cm}
\includegraphics[width=2.9in]{GA-np-M} }
\end{center}
\vspace{.1cm}
\begin{center}
\mbox{
\includegraphics[width=2.9in]{GA-XmX0B} \hspace{1.cm}
\includegraphics[width=2.9in]{GA-XmX0-M1} }
\end{center}
\caption{\footnotesize{Axial-vector form factor $G_A$ for
    $|\Delta I| =1$ transitions.
    {\bf Left panel:} results in vacuum (bare, meson cloud and total).
    {\bf Right panel:} total results for the medium $\rho=0.5 \rho_0$ and
    $\rho_0$ compared with vacuum ($\rho=0$).
    The results for the $\Sigma^+ \to \Lambda$ and
    $\Sigma^0 \to \Sigma^+$ transitions are
    presented in Appendix~\ref{app-GA}, Fig.~\ref{figGA-p1X}.
}}
\label{figGA-p1}
\end{figure*}

\begin{figure*}[t]    %%%%    Figure 4 (12)    UPDATE
\vspace{.5cm}
\begin{center}
\mbox{
\includegraphics[width=2.9in]{GA-Lp} \hspace{1.cm}
\includegraphics[width=2.9in]{GA-Lp-M}}
\end{center}
 \vspace{.1cm}
 \begin{center}
\mbox{
\includegraphics[width=2.9in]{GA-S0p} \hspace{1.cm}
\includegraphics[width=2.9in]{GA-S0p-M} }
 \end{center}
  \vspace{.1cm}
 \begin{center}
\mbox{
\includegraphics[width=2.9in]{GA-X0SpB} \hspace{1.cm}
\includegraphics[width=2.9in]{GA-X0Sp-M1} }
\end{center}
\caption{\footnotesize{Axial-vector form factor $G_A$
for $|\Delta S| =1$ transitions.
    {\bf Left panel:} results in vacuum (bare, meson cloud and total).
    {\bf Right panel:} total results for the medium $\rho=0.5 \rho_0$ and
    $\rho_0$ compared with vacuum ($\rho=0$).
    The results for the $\Sigma^- \to n$ transition are
    presented in Appendix~\ref{app-GA}, Fig.~\ref{figGA-p21X}.
    The results for $\Xi^- \to \Lambda$ and $\Xi^- \to \Sigma^0$
    transitions are
    presented in Appendix~\ref{app-GA}, Fig.~\ref{figGA-p22X}.   }}
\label{figGA-p21}
\end{figure*}

We use the ratio between the form factors in medium and
vacuum for a clear visualization of the medium modifications (Sec.~\ref{sec-Medium2}).
Deviations from unity (one) are the signature of medium effects.
We study also the slope of the form factors near $Q^2=0$.

We summarize the results for $G_A$ and $G_P$ in medium in Sec.~\ref{sec-Summary1}.
We finish the section with a general discussion
of our results, the limitations of the model, 
and the possible improvements (Sec.~\ref{sec-discussion}).

%%%%%%%%%%%%%%%%%%%%%%%%%%%%%%%%%%%%%%%%%%%%%%%%%%%%%%%%%%%%%%%%%%%%%%
%%%%%     FIGURES
%%%%%%%%%%%%%%%%%%%%%%%%%%%%%%%%%%%%%%%%%%%%%%%%%%%%%%%%%%%%%%%%%%%%%%

\begin{figure*}[t]    %%   Fig. 6  (14)
  %\vspace{.5cm}
\begin{center}
\mbox{
\includegraphics[width=2.9in]{GP-np} \hspace{1.cm}
\includegraphics[width=2.9in]{GP-np-M} }
\end{center}
\vspace{.1cm}
\begin{center}
 \mbox{
\includegraphics[width=2.9in]{GP-XmX0} \hspace{1.cm}
\includegraphics[width=2.9in]{GP-XmX0-M} }
\end{center}
\caption{\footnotesize{Induced pseudoscalar form factor $G_P$ for $|\Delta I|=1$ transitions.
    {\bf Left panel:} results for bare, bare plus meson cloud and total.
    {\bf Right panel:} total results for the medium $\rho=0.5 \rho_0$ and
    $\rho_0$ compared with vacuum ($\rho=0$).
The results for the $\Sigma^+ \to \Lambda$ and
    $\Sigma^0 \to \Sigma^+$ transitions are
    presented in Appendix~\ref{app-GP}, Fig.~\ref{figGP-p1X}.
   }}
\label{figGP-p1}
\end{figure*}

\begin{figure*}[t]    %%%%   Figure 7  (page 15)  ===>    Fig. 6
  %\vspace{.5cm}
\begin{center}
\mbox{
\includegraphics[width=2.9in]{GP-Lp} \hspace{1.cm}
\includegraphics[width=2.9in]{GP-Lp-M}}
\end{center}
 \vspace{.1cm}
 \begin{center}
\mbox{
\includegraphics[width=2.9in]{GP-S0p} \hspace{1.cm}
\includegraphics[width=2.9in]{GP-S0p-M} }
 \end{center}
  \vspace{.16cm}
 \begin{center}
\mbox{
\includegraphics[width=2.9in]{GP-X0Sp} \hspace{1.cm}
\includegraphics[width=2.9in]{GP-X0Sp-M} }
\end{center}
 \caption{\footnotesize{Induced pseudoscalar form factor $G_P$
     for $|\Delta S| =1$ transitions.
     {\bf Left panel:} results for bare, bare plus meson cloud and total.
{\bf Right panel:} total results for the medium $\rho=0.5 \rho_0$ and
$\rho_0$ compared with vacuum ($\rho=0$).
    The results for the $\Sigma^- \to n$ transition are
    presented in Appendix~\ref{app-GP}, Fig.~\ref{figGP-p21X}.
    The results for $\Xi^- \to \Lambda$ and $\Xi^- \to \Sigma^0$
    transitions are
    presented in Appendix~\ref{app-GP}, Fig.~\ref{figGP-p22X}. }}
\label{figGP-p21}
\end{figure*}

%\newpage
%\clearpage

\subsection{Axial form factors in vacuum \label{sec-Vacuum}}

Using the formalism given in Sec.~\ref{secVacuum},
we calculate the form factors $G_A$ and $G_P$ in vacuum.
These calculations are the extension of the parametrization for the 
nucleon (Fig.~\ref{figGAGP-Nucleon}) 
to the octet baryons.
This extension is based on our analysis of the nucleon lattice and physical data,
combining the bare contributions with
an effective parametrization of the meson cloud.
The meson cloud component is calibrated by the
nucleon physical data and for the octet baryon  $G_A(0)$ physical data.

%%   XVSPACE

%\vspace{.3cm}

%\vspace{1.cm}

\subsubsection*{Results for $G_A$}

Some examples of the $B' \to B$ octet baryon axial-vector form factors $G_A$
are presented on the left side in 
Figs.~\ref{figGA-p1}  ($|\Delta I| =1$) and \ref{figGA-p21} ($|\Delta S| =1$).
The results for all transitions are presented in Appendix~\ref{app-GA}
(Figs.~\ref{figGA-p1X}, \ref{figGA-p21X} and \ref{figGA-p22X}).
The contributions from the quark core are represented by the
dash lines and the contributions from the meson cloud by
the dash-doted lines.
The final results correspond to the solid lines.
In the cases where the final result $G_A$ is negative,
we represent the function $-G_A$, for an easier comparison
with the other reactions.
In the figures we include also the available data for $G_A(0)$.

From these figures and from Figs.~\ref{figGA-p1X}, \ref{figGA-p21X} and \ref{figGA-p22X},
we can conclude that we can have
different relative magnitudes for the form factor,
for the bare contribution and for the meson cloud contribution.
Most transitions are dominated by the bare contributions
but some transitions have significant relative meson cloud contributions
($\Xi^- \to \Xi^0$, $\Sigma^- \to n$, $\Sigma^0 \to p$ and $\Xi^0 \to \Sigma^+$).
Transitions with a large magnitude ($|G_A(0)|   \gtrsim 1$)
can have about 33\% of meson cloud contributions
(transitions $n \to p$, $\Lambda \to p$,
$\Xi^- \to \Sigma^0$ and $\Xi^0 \to \Sigma^+$).

In comparison with previous work~\cite{AxialFF}
we correct here the normalization of the $\Xi$ baryons, relatively
to the previous work, as discussed already in Sec.~\ref{secVacuum}.
The main difference to the previous calculations 
are then the results for $\Xi^- \to \Lambda$ and $\Xi^0 \to \Sigma^+$,
which are modified by the value of $Z_\Xi$.
As a consequence, both, the bare and the 
meson cloud contributions are enhanced.
With the present parametrization
the description of $\Xi^- \to \Lambda$ transition is improved,
while the numerical result for $\Xi^0 \to \Sigma^+$ overestimates the data.

Since the estimates follow approximate SU(6) and SU(3) symmetries,
and the results for $\Xi^0 \to \Sigma^+$ are expected to be similar 
to the results for $n \to p$, we may question the quality 
of the SU(3) description for the meson cloud component.
We note, however, that these results are a
consequence of the normalization of the $\Xi$ states,
based on the studies of the electromagnetic structure 
of the octet baryons,  which assume the dominance of the
pion cloud for the meson cloud.
This leads to a stronger suppression in the nucleon system ($Z_N \simeq 0.73$)
than in the $\Xi$ systems ($Z_\Xi \simeq 0.98$).
The inclusion of the kaon cloud for the systems $\Lambda$, $\Sigma$
and $\Xi$ (expected to be more significant) imply a correction of 
the denominator in Eq.~(\ref{eqZB1})
and a reduction of the factor $Z_B$ for the systems with one or
two strange quarks.
As a consequence the estimates for the $\Xi$ axial form factors
would be reduced, improving the description of the data.

%%%%%%%%%%%%%%%%%%%%%%%%%%%%%%%%%%%%%%%%%%%%%%%%%%%%%%%%%%%%%%%%%%%%%

\begin{figure*}[t]   %%    Fig. 9  (17)
  %\vspace{.5cm}
\begin{center}
\mbox{
\includegraphics[width=2.9in]{GA-np-RT} \hspace{1.cm}
\includegraphics[width=2.9in]{GA-np-R} }
\end{center}
 \vspace{.1cm}
%\centerline{ \vspace{.1cm}}
 %\vspace{.1cm}
\begin{center}
\mbox{
\includegraphics[width=2.9in]{GA-XmX0-RTB} \hspace{1.cm}
\includegraphics[width=2.9in]{GA-XmX0-RB}}
\end{center}
\caption{\footnotesize{Axial-vector form factor $G_A$ for $|\Delta I|=1$ transitions.
    Ratios $G_A^\ast/G_A$ and $G_A^\ast(Q^2)/G_A^\ast(0)$.
    The results for the $\Sigma^+ \to \Lambda$ and
    $\Sigma^0 \to \Sigma^+$ transitions are
    presented in Appendix~\ref{app-GA}, Fig.~\ref{figGA-p30X}.
}}
\label{figGA-p30}
\end{figure*}

\begin{figure*}[t]   %%   Fig.10   (19)
  %\vspace{.5cm}
\begin{center}
\mbox{
\includegraphics[width=2.9in]{GA-Lp-RT} \hspace{1.cm}
\includegraphics[width=2.9in]{GA-Lp-R}}
\end{center}
%\centerline{ \vspace{.05cm}}
\vspace{.1cm}
 \begin{center}
\mbox{
\includegraphics[width=2.9in]{GA-X0Sp-RTB} \hspace{1.cm}
\includegraphics[width=2.9in]{GA-X0Sp-RB} }  %GA-X0Sp-RB
\end{center}
\caption{\footnotesize{Axial-vector form factor $G_A$ for $|\Delta S| =1$ transitions.
    Ratios $G_A^\ast/G_A$ and $G_A^\ast(Q^2)/G_A^\ast(0)$.
    The results for the $\Sigma^- \to n$, $\Sigma^0 \to p$
      and $\Xi^- \to \Lambda$
      transition are presented in Appendix~\ref{app-GA}, Fig.~\ref{figGA-p31X}.
    The ratios are the same for $\Sigma^- \to n$ and $\Sigma^0 \to p$.
    The ratios are the same for  $\Xi^- \to \Sigma^0$ and $\Xi^0 \to \Sigma^+$.
}}
\label{figGA-p31}
\end{figure*}

%%%%     Figures 9, 10
%%%%      FIGURES    12, 12

%%   XVSPACE

%\newpage

\subsubsection*{Results for $G_P$}

Some examples of our predictions for $G_P$ are on the left side of
Fig.~\ref{figGP-p1} for $|\Delta I|=1$,
and on the left side of Fig.~\ref{figGP-p21} for $|\Delta S| =1$.
The results for $G_P$ for all transitions are presented
in Appendix~\ref{app-GP} (Figs.~\ref{figGP-p1X}, \ref{figGP-p21X} and \ref{figGP-p22X}).
As for the function $G_A$, in the cases where the final result $G_P$ is negative,
we represent the function $-G_P$.
In the figures, we discriminate the result for the bare contribution,
the combination of the bare and pseudoscalar meson pole contribution,
and the sum of all the contributions (Bare + Pole + Meson Cloud).
We adjust the upper limits of $Q^2$
in order to discriminate the different contributions at low $Q^2$
according to the falloff of the form factors.

In the calculations for $|\Delta I| =1$
(Figs.~\ref{figGP-p1} and \ref{figGP-p1X}) it is clear the 
impact of the factor $1/(m_\pi^2 + Q^2)$ 
in the Bare + Pole result and in the final result.
The strong reduction with $Q^2$ is a consequence of
the small magnitude of $m_\pi$.
In the case of the nucleon,
the combination of the bare contribution and
the meson cloud contribution is decisive for the 
good agreement of the model with the physical form factor data
(see Fig.~\ref{figGAGP-Nucleon} and Sec.~\ref{secVacuum}).
In the remaining cases our calculations are pure predictions.

The results for $|\Delta S|=1$ (Figs.~\ref{figGP-p21}, \ref{figGP-p21X} and \ref{figGP-p22X})
show a slower falloff for the form factor $G_P$ than for $|\Delta I| =1$,
as expected from the kaon pole contribution associated with the factor $1/(m_K^2 + Q^2)$.
The magnitude of $G_P$ decreases slower with $Q^2$
than for $|\Delta I| =1$ as a consequence of the kaon mass of the pole.
The differences of magnitudes near $Q^2=0$ in comparison with $|\Delta I| =1$  
are mainly the consequence of the amplification of the meson mass.

Another important point is the relative magnitude of the terms.
The bare and pole terms have, in the case  $|\Delta S| =1$, similar magnitudes,
and there is a partial cancellation between the pole terms
and the bare terms, which have opposite signs for all transitions.
Notice that, since the pole and meson cloud contributions
have the same sign for the transitions under discussion,
one can also say that there is a partial cancellation
between bare and meson cloud contributions at low $Q^2$.
The consequence of these effects is that the form factors $G_P$ for $|\Delta S| =1$
are about one order of magnitude smaller than the
form factors $G_P$ for $|\Delta I|=1$.
In a few cases ($\Xi^- \to \Lambda$, $\Xi^- \to \Sigma^0$ and $\Xi^0 \to \Sigma^+$)
the bare contribution has a magnitude comparable with
the meson cloud contribution at \mbox{low $Q^2$.}
%low $Q^2$.

For future discussion, notice that, near the upper limit of $Q^2$
one can observe the convergence of the full result (solid line)
and the Bare + Pole contribution (dashed line).
This result means that in the region where the bare contribution
is reduced (close to the horizontal zero line) the meson cloud contributions
become also negligible.
Recall that for large $Q^2$, we expect $G_P^{\rm MC} \propto 1/Q^{10}$.
Notice also that, although, $G_P^{\rm B}$ and $G_P^{\rm Pole}$
are expected to have similar asymptotic behaviors ($\propto 1/Q^6$),
numerically $G_P^{\rm B}$ falls down much faster,
since $G_P \simeq G_P^{\rm B} + G_P^{\rm pole} \simeq G_P^{\rm pole}$ for large $Q^2$.
The small magnitude of $G_P^{\rm B}$ at large $Q^2$
is a consequence of the model calibration
of the $G_P$ lattice QCD data for the nucleon.

Our last remark about $G_P$ is the enhancement of $G_P$ near $Q^2=0$
as a consequence of the low-$Q^2$ turning point of the function $G_P^{\rm B}$.
The effect appears also for  $|\Delta I| =1$ (Figs.~\ref{figGP-p1} and \ref{figGP-p1X}),
but it is overshadowed by
the magnitude of the pole and meson cloud terms.
It can be observed more clearly in the figures for $|\Delta S| =1$
(Figs.~\ref{figGP-p21}, \ref{figGP-p21X} and \ref{figGP-p22X}).
This behavior is a consequence of the method used
to calibrate the model: an unconstrained fit to the 
lattice QCD data for the nucleon.
The lattice datasets used in the calibration include no $G_P$
data below $Q^2=0.15$ GeV$^2$~\cite{Alexandrou11a}.
Therefore, it is possible that
$G_P^{\rm B}$ may be poorly constrained, affecting consequently
the accuracy of the description of $G_P$ near $Q^2=0$.

In the present study, we use the calibration from Ref.~\cite{AxialFF}
as a first approximation.
The results are, in principle, more accurate for the $|\Delta I| =1$ sector,
less sensitive to the magnitude of the function $G_P^{\rm B}$.
Future improvements of the present scheme are discussed
in Sec.~\ref{sec-discussion}.

\subsection{Axial form factors in medium  \label{sec-Medium1}}

Some examples of our extension of the model 
to the octet baryons in the nuclear medium
are presented on the right side
of Figs.~\ref{figGA-p1} ($|\Delta I| =1$) and 
\ref{figGA-p21} ($|\Delta S| =1$) for $G_A$, 
and on the right side of Figs.~\ref{figGP-p1}
 ($|\Delta I| =1$) and 
\ref{figGP-p21} ($|\Delta S| =1$) for $G_P$.
The results for all transitions are presented in Appendix~\ref{app-GA}
(Figs.~\ref{figGA-p1X}, \ref{figGA-p21X} and \ref{figGA-p22X}) for $G_A$,
and Appendix~\ref{app-GP}
(Figs.~\ref{figGP-p1X}, \ref{figGP-p21X} and \ref{figGP-p22X}) for $G_P$.

For the nuclear medium, for simplicity, we present only the final results for the
densities $\rho=0.5 \rho_0$ and $\rho=\rho_0$,
in comparisons with the result in vacuum ($\rho=0$),
omitting the decomposition of the bare and meson cloud contributions.

For the discussion of the results in medium
it is important to look to the ratios $G_A^\ast/G_A$
presented on the left side of Figs.~\ref{figGA-p30} 
and \ref{figGA-p31} and the ratios $G_P^\ast/G_P$
presented on the left side of Figs.~\ref{figGP-p30} and \ref{figGP-p31}.
The ratios for all transitions are presented
in Appendix~\ref{app-GA} (Figs.~\ref{figGA-p30X} and \ref{figGA-p31X}) for $G_A$,
and in Appendix~\ref{app-GP} (Figs.~\ref{figGP-p30X} and \ref{figGP-p31X})  for $G_P$.
In this respect, we recall the discussion in Sec.~\ref{secBpMC},
according to which the ratios for  $\Sigma^0 \to p$ are the same
as the ratios for $\Sigma^- \to n$.
The same property is verified for the
transitions $\Xi^0 \to \Sigma^+$ and $\Xi^- \to \Sigma^0$.

%%   HVSPACE

%\newpage

\hspace{.25cm}

\subsubsection{Results for $G_A^\ast$ \label{secGA-star}}

From the results for the form factor $G_A$ in medium:
right side of Figs.~\ref{figGA-p1}, \ref{figGA-p21},
\ref{figGA-p1X}, \ref{figGA-p21X} and \ref{figGA-p22X}
for the function $G_A^\ast$,
and left side of Figs.~\ref{figGA-p30}, \ref{figGA-p31}, \ref{figGA-p30X} and \ref{figGA-p31X}
for the function $G_A^\ast/G_A$,
we conclude that, in general, the axial-vector form factor $G_A$
is reduced in the nuclear medium, or symmetric nuclear matter 
(suppression of $G_A$ in medium).
The conclusion is valid for $|\Delta I| =1$ and $|\Delta S| =1$.

The quenching of $G_A$ for the octet baryon is mainly
a consequence of the effect discussed already in the case
of the nucleon (see Sec.~\ref{secMedium}).
The enhancement of the $P$-state components in medium
leads to reduction of the functions compared to vacuum.

In the present case these results are a consequence of the suppression in medium
of both the meson cloud and bare contributions.
The suppression of the meson cloud contribution is more effective at low $Q^2$,
while the suppression of the bare contributions is more effective at large $Q^2$.
The magnitude of the suppression can be
better observed using the function $G_A^\ast/G_A$
in terms of $Q^2$
(left side of Figs.~\ref{figGA-p30} and \ref{figGA-p30X} for $|\Delta I| =1$,
and left side of Figs.~\ref{figGA-p31} and \ref{figGA-p31X} for $|\Delta S| =1$).
The suppression is stronger for lighter baryons
and larger densities.

%% XVSPACE

%\newpage

\vspace{.7cm}

From the observation of Figs.~\ref{figGA-p30}, \ref{figGA-p31},
\ref{figGA-p30X} and \ref{figGA-p31X}, we can classify the ratios $G_A^\ast/G_A$ using the shape and 
the magnitude of the suppression in the order:
\begin{itemize}
\item
  $n \to p$,
\item
  $\Lambda \to p$, $\Sigma^+ \to p$/$\Sigma^- \to n$,
\item
  $\Sigma^+ \to \Lambda$, $\Sigma^0 \to \Sigma^+$, $\Xi^- \to \Lambda$,
  $\Xi^- \to \Sigma^0$/$\Xi^0 \to \Sigma^+$,
\item
  $\Xi^- \to \Xi^0$.
\end{itemize}
The range of variation discussed next is
relative to the interval $Q^2=0$--2 GeV$^2$.

The first reaction ($n \to p$) includes no transitions
between the strange and light quarks,
and, thus, it is the reaction with stronger suppression in medium.
The effect increases with $Q^2$.
For $\rho/\rho_0 =0.5$
one has a reduction of about 5--18\%.
For $\rho/\rho_0 =1.0$ the reduction varies from 8--32\%.

The second type includes the $\Lambda \to p$ transition,
that is similar to the $n \to p$ transition, but with a weaker reduction
of the ratio $G_A^\ast/G_A$.
The $\Sigma^+ \to p$/$\Sigma^- \to n$ ratio has a similar shape.

The third class of transitions
($\Sigma \to \Lambda, \Sigma$,
$\Xi \to \Lambda, \Sigma$) includes systems composed by a strange quark
in the final state and have smaller and slower
reductions with $Q^2$.

The last case ($\Xi^- \to \Xi^0$) is a transition between
baryons with two strange quarks and reveals a significant suppression
at low $Q^2$ (meson cloud contributions significantly reduced in medium)
and only a mild dependence on $Q^2$, for $Q^2 > 1$ GeV$^2$.

Our estimates for $G_A^\ast/G_A$ for the nucleon can be compared with
the calculations of the bag model~\cite{Lu01a,Cheoun13a,Cheoun13b}.
The main difference between our calculations and
the bag model calculations is that the latter
does not include the effects of the meson cloud dressing.
Apart from small differences at low $Q^2$, the calculations
differ on the shape of the ratio.
The bag model ratio estimates decrease till the region $Q^2=0.5$--1 GeV$^2$,
and start to increase after that. 
After a certain value of $Q^2$, one has an enhancement of $G_A^\ast$.
This is due to the Lorentz contraction effect that is taken into account
for the ''bag shape'' in Ref.~\cite{Lu01a}, and,  thus, for larger $Q^2$
the effect starts to appear stronger.
Or, this is due to the finite size effect of the nucleon bag which is subject to
the Lorentz contraction.
The present calculations in contrast, suggest that the medium suppression
keeps increasing for large $Q^2$ although with a reduced rate.

%%   HVSPACE

%\newpage

\subsubsection{Results for $G_P^\ast$ \label{secGP-star}}

Examples of our calculations for the induced pseudoscalar form factor $G_P$
in the nuclear medium are presented on the right side of
Fig.~\ref{figGP-p1} for $|\Delta I| =1$ and 
Fig.~\ref{figGP-p21}  for $|\Delta S| =1$.
The results for all transitions are presented in 
Appendix~\ref{app-GP} (Figs.~\ref{figGP-p1X}, \ref{figGP-p21X} and \ref{figGP-p22X}).
Examples of ratios $G_P^\ast/G_P$
are presented on the left side of Figs.~\ref{figGP-p30}
for $|\Delta I|=1$ and \ref{figGP-p31} for $|\Delta S| =1$.
The comparison of the ratios for all transitions
is presented  in Appendix~\ref{app-GP} (Figs.~\ref{figGP-p30X} and \ref{figGP-p31X}).
The range of $Q^2$ is adjusted to each case
in order to better visualize the shape of the functions.

On the right side of
Figs.~\ref{figGP-p1}, \ref{figGP-p21}, \ref{figGP-p1X}, \ref{figGP-p21X} and \ref{figGP-p22X},
we can still observe
the enhancement of $G_P$ near $Q^2=0$ as in vacuum,
for both $|\Delta I| =1$ and $|\Delta S| =1$.
Since, as mentioned already, the $G_P$ estimates at $Q^2=0$
may be inaccurate, we avoid definitive conclusions
about the point $Q^2=0$, and
for the ratios $G_P^\ast/G_P$ in Figs.~\ref{figGP-p30}, 
\ref{figGP-p31}, \ref{figGP-p30X} and \ref{figGP-p31X},
we display only the region $Q^2 > 0.012$ GeV$^2$.

In general, one can observe an overall suppression
of the form factors $G_P^\ast$ for both cases
($|\Delta I| =1$ and $|\Delta S| =1$)
compared with the vacuum,
particularly at low $Q^2$. 
The suppression increases with the density.
The suppression at low $Q^2$ is diluted when $Q^2$ increases.
In some cases, there are apparent overlaps
of the  different calculations
($\rho=0$, $0.5 \rho_0$ and $\rho_0$),
particularly for heavy baryons.
See the last two lines on the right side of  
Fig.~\ref{figGP-p1X}
and all lines on the right side of Fig.~\ref{figGP-p22X}.
The apparent overlaps are the consequence of 
the reduced scale in $Q^2$ and the thickness of the lines.

The analysis of the suppression effect in medium on $G_P$
is simplified when we look at the ratios $G_P^\ast/G_P$
that are presented on the left side of Figs.~\ref{figGP-p30} and \ref{figGP-p30X}
for $|\Delta I| =1$ and on the left side of Figs.~\ref{figGP-p31} and \ref{figGP-p31X}
for $|\Delta S| =1$.

In some cases ($\Sigma^+ \to \Lambda$, $\Sigma^0 \to \Sigma^+$, 
$\Xi^- \to \Lambda$ and $\Xi^- \to \Sigma^+$)
we have more significant reductions of $G_P^\ast/G_P$ at low $Q^2$
due to the more effective reduction
of the meson cloud component, as discussed already for $G_A^\ast$
(recall that $G_P^\ast  \propto  G_A^\ast$).
At large $Q^2$, we can expect for sufficient large values
of $Q^2$ the dominance of the pole term (proportional to $G_A^{{\rm B} \ast}$).
This effect can be observed in the figures for
the vacuum
(left side of Figs.~\ref{figGP-p1X}, \ref{figGP-p21X} and \ref{figGP-p22X})
in the upper limit of $Q^2$.

We start the discussion with the  $|\Delta I| =1$ transitions. 
For the transitions $|\Delta I| =1$, on the left side of Figs.~\ref{figGP-p30}
and \ref{figGP-p30X},
the significant reduction of the ratio $G_P^\ast /G_P$ can
be understood as a consequence of the dominance of the
pole and meson cloud terms, a sum proportional to
$G_A^\ast = G_A^{{\rm B} \ast} + G_A^{{\rm MC} \ast}$.
In that case, $G_P^\ast \propto (M_{BB'}^{\ast})^2 G_A^\ast $, since
in the present model $m_\pi$ has the same value
for the vacuum and medium
($m_\pi$ is almost insensitive to the medium
modifications~\cite{Hayano,Kienle,Meissner,Vogl}).

In these conditions, we can conclude that
\ba
\frac{G_P^\ast}{G_P}  \simeq
\left( \frac{M_{B}^\ast + M_{B'}^\ast}{M_{B} + M_{B'}} \right)^2 \frac{G_A^\ast}{G_A}.
\label{eqGPmedium}
\ea

The result $G_P^\ast/G_P < 1$ is then naturally explained
by the mass reduction in medium
($\frac{M_{B}^\ast + M_{B'}^\ast}{M_{B} + M_{B'}} < 1$)
and by the reduction of $G_A$ in medium ($G_A^\ast/G_A < 1$)
discussed already.
The suppression is more effective due to
the power 2 in the mass ratio.
One can then expect a triple suppression
of the ratio $G_P^\ast/G_P$ in medium for $|\Delta I| =1$.
The impact of the suppression is smaller
for heavy baryons
(mass ratio and $G_A^\ast/G_A$ are reduced more weakly),
and lower densities $\rho$.
Note, in this respect, that the reduction of $G_P^\ast/G_P$
is very mild for the $\Xi^- \to \Xi^0$ near the upper limit of $Q^2$.

We consider now the $|\Delta S| = 1$ transitions.
In this case, the pole and the meson cloud term
include the factor $\frac{(M_B^\ast + M_{B'}^\ast)^2}{m_{K}^{* 2} + Q^2}$.
At low $Q^2$ the effect of mass reduction of the kaon
cancels in part the effect of the reduction of the baryon masses in medium
(factor $\frac{(M_B^\ast + M_{B'}^\ast)^2}{(M_B + M_{B'})^2}$),
since $\frac{m_K^2}{m_{K}^{\ast 2}} > 1$.
Thus, at low $Q^2$, the ratio is dominated by the
reduction of the meson cloud contribution
(more significant reduction at low $Q^2$).
For large $Q^2$, the effect of the kaon mass
modification $m_K \to m_K^\ast$
is included in the factor
$1 + \frac{m_K^2 - m_K^{* 2}}{ m_K^{* 2} + Q^2}$ and 
decreases quickly to the unit.
The consequence is that ratio $G_P^\ast/G_P$ 
can be described by Eq.~(\ref{eqGPmedium}),
with $G_A^\ast$ replaced by $G_A^{B \ast}$
(dominance of the pole term at large $Q^2$). 
The conclusion is then that at large $Q^2$,
we can expect a stronger suppression for light baryons
(more significant baryon mass effect and
reduction of $G_A^{{\rm B}\ast}/G_A^{\rm B}$)
and almost no medium effects for heavy baryons ($G_P^\ast/G_P \approx 1$).

The final conclusion is then that $G_P^\ast/G_P < 1$
if the mass reduction effect is significant (light baryons, large densities).
For heavy baryons, the mass reduction is less effective, 
and we can observe the convergence of the calculation
($G_P^\ast/G_P \approx 1$) for relatively small values of $Q^2$
(see transitions $\Xi^- \to \Lambda$,  $\Xi^- \to \Sigma^0$ 
and   $\Xi^0 \to \Sigma^+$). 

%% VSPACE

\subsection{Axial form factors in medium --
  Falloffs with $Q^2$  \label{sec-Medium2}}

To finish the discussion about the form factors
$G_A$ and $G_P$ in medium we analyze the dependence of the form factors on $Q^2$.
This information is important when we discuss the
slope of the form factors near $Q^2=0$,
at large $Q^2$, or parametrize the form factors (vacuum or medium)
in terms of simple dipole forms.

Notice that we do not discuss
now the variation $G_A^\ast$ relative to $G_A$
in terms of  $Q^2$ (the same for $G_P^\ast$ and $G_P$),
but that we are now interested in the variation
of the form factors relative to the result at $Q^2=0$.
The ratios discussed in the previous section give
us information about the amount (percentage) of
variation from the vacuum to the medium, but tell us nothing
about the relative falloff of the form factors.

We look now at the functions
$G_A^\ast(Q^2)/G_A^\ast(0)$ and $G_P^\ast(Q^2)/G_P^\ast(0)$.
In the limit $Q^2=0$ the slope of the ratio
gives information about the derivatives of the functions at $Q^2=0$.
Taking $G_A$ as an example, one has for small $Q^2$:
$G_A(Q^2) - G_A(0)= - \frac{1}{6}\left< r_A^2\right> Q^2 G_A(0)$,
where the axial-vector square radius is
$\left< r_A^2\right>= -\frac{6}{G_A(0)} \frac{d G_A}{d Q^2}(0)$.
A similar form can be used for $G_P$.

The ratios for finite $Q^2$ measure the variation
of the falloff with $Q^2$ and can be used to observe variations
from simple dipole forms.

%%%%%%%%%%%%%%%%%%%%%%%%%%%%%%%%%%%%%%%%%%%%%%%%%%%%%%%%%%%%%%%%%%%%%%%%%%%%%%%%%%%%%%%%
%                        FIGURES   --   PART 2
%%%%%%%%%%%%%%%%%%%%%%%%%%%%%%%%%%%%%%%%%%%%%%%%%%%%%%%%%%%%%%%%%%%%%%%%%%%%%%%%%%%%%%%%

%%%%%%%%%%%%%%%%%%%%%%%%%%%%%%%%%%%%%%%%%%%%%%%%%%%%%%%%%%%%%%%%%%%%
%
%    FORM FACTORS RATIOS    GP*(Q2)/GP*(0)
%
%%%%%%%%%%%%%%%%%%%%%%%%%%%%%%%%%%%%%%%%%%%%%%%%%%%%%%%%%%%%%%%%%%%%

\begin{figure*}[t]      %%  Fig. 11  (21)
  %\vspace{.5cm}   %%  vspace can be avoided
\begin{center}
\mbox{
\includegraphics[width=2.8in]{GP-np-RT} \hspace{1.cm}
\includegraphics[width=2.9in]{GP-np-R} }
\end{center}
\vspace{.13cm}
\begin{center}
\mbox{
\includegraphics[width=2.8in]{GP-XmX0-RTB} \hspace{1.cm}
\includegraphics[width=2.9in]{GP-XmX0-RB}}
\end{center}
\caption{\footnotesize{Induced pseudoscalar form factor
    $G_P$ for $|\Delta I| =1$ transitions.
    Ratios $G_P^\ast/G_P$ and $G_P^\ast(Q^2)/G_P^\ast(Q_p^2)$.
    The results for the $\Sigma^+ \to \Lambda$ and
    $\Sigma^0 \to \Sigma^+$ transitions are
    presented in Appendix~\ref{app-GP}, Fig.~\ref{figGP-p30X}.
}}
\label{figGP-p30}
\end{figure*}

\begin{figure*}[t]    %%  Fig. 12  (23)
  %\vspace{.5cm}
\begin{center}
\mbox{
\includegraphics[width=2.8in]{GP-Lp-RT} \hspace{1.cm}
\includegraphics[width=2.9in]{GP-Lp-R}}
\end{center}
\vspace{.15cm}
 \begin{center}
\mbox{
\includegraphics[width=2.9in]{GP-X0Sp-RTB} \hspace{1.cm}
\includegraphics[width=2.9in]{GP-X0Sp-RB} }
\end{center}
\caption{\footnotesize{Induced pseudoscalar form factor $G_P$ for $|\Delta S| =1$ transitions.
    Ratios $G_P^\ast/G_P$ and $G_P^\ast(Q^2)/G_P^\ast(Q_p^2)$.
      The results for the $\Sigma^- \to n$, $\Sigma^0 \to p$
      and $\Xi^- \to \Lambda$
      transition are presented in Appendix~\ref{app-GP}, Fig.~\ref{figGP-p31X}.
The ratios are the same for  $\Xi^- \to \Sigma^0$ and $\Xi^0 \to \Sigma^+$.
  }}
\label{figGP-p31}
\end{figure*}

\subsubsection{Axial-vector form factor $G_A^\ast$}

Examples of calculations of $G_A^\ast(Q^2)/G_A^\ast(0)$
are presented on the right side of Figs.~\ref{figGA-p30} and~\ref{figGA-p31},
for $|\Delta I| =1$ and $|\Delta S| =1$, respectively.
The results for all transitions are presented in Appendix~\ref{app-GA}
(Figs.~\ref{figGA-p30X} and \ref{figGA-p31X}).

The results can be divided into three main cases
in terms of the dependence on $\rho$:
(i) $n \to p$ and $\Lambda \to p$ transitions:
differences for large $Q^2$; (ii) $\Sigma^- \to n$ transition:
almost no dependence;
(iii) remaining cases: differences at low and large $Q^2$.
Even in the last case, the dependence on $\rho$ is not strong
(variation of a few percent).
The main conclusion is then that the effect of the medium
is more significant in the magnitude of $G_A^\ast$,
discussed in the previous section.

As mentioned already, the slope of $G_A^\ast(Q^2)/G_A^\ast(0)$
is related to the axial-vector square radius 
$\left< r_A^2\right>$ in the limit $Q^2=0$.
The results displayed on Figs.~\ref{figGA-p30}, \ref{figGA-p31},
\ref{figGA-p30X}, and \ref{figGA-p31X} suggest,
however, that the slope is smaller than
that we can expect from a typical dipole form (\ref{eqGAexp})
with $M_A \simeq 1.05$ GeV:
$\left< r_A^2\right> = \frac{12}{M_A^2} \simeq 0.42$ fm$^2$ 
for the $n \to p$ transition.
The axial-vector square radius obtained for
the vacuum is of the order of 0.1 fm$^2$ for the 
$n \to p$, $\Sigma^0 \to \Sigma^+$ transitions;
0.16 fm$^2$ for $\Xi^- \to \Sigma^0$ and $\Xi^0 \to \Sigma^+$
transitions; 0.2 fm$^2$ for the $\Sigma^+ \to \Lambda$
and about 0.35 fm$^2$ for the $\Xi^- \to \Xi^0$,
$\Lambda \to p $, $\Sigma^- \to n$ and $\Sigma^0 \to p$.
The $\Xi^- \to \Lambda$ transition is more atypical
and has a negative axial-vector square radius
(as can be inferred from the figure).
These results are a consequence of the smooth dependence
of the function $G_A$ near $Q^2=0$, that can be linked with 
$P$-state contributions to the transition form factors.
This topic is discussed in Sec.~\ref{sec-discussion}.
The conclusion is then that near $Q^2=0$, at the scale
of $Q^2=0.01$ GeV$^2$ the axial-vector form factor $G_A$
has a slower falloff than suggested by a
dipole parametrization of $G_A$ with a typical cutoff $M_A \simeq 1.05$ GeV,
for the $n \to p$ transition
(0.13 fm$^2$ to be compared with the {\it experimental} estimate
0.42 fm$^2$).

The present results suggest then that $\left< r_A^2\right>$
may decrease slightly in medium,
contrary to what we naively observe, for instance by the calculations
of the corresponding electromagnetic form factors,
namely, charge radii of the
octet baryons increase in medium~\cite{Octet2,Octet3,JLabbook,QMCEMFFMedium}.
In contrast, the calculations of  $\left< r_A^2\right>$ from Refs.~\cite{Petti23a,Kaiser24a}
suggest that we may expect a small enhancement in medium.

We notice, however, that the slow falloff of $G_A$
near $Q^2=0$ is a local effect, and when we
extend the analysis to the range $Q^2=0.1$ GeV$^2$,
our model calculation for $G_A$ compares well with
the dipole parametrization discussed above.
Only for $Q^2 > 1.2$ GeV$^2$, do we observe significant differences
from the dipole parametrization.
Our model has a falloff slower than the dipole parametrization.

The conclusion of this section is that
(i) the falloff
does not vary significantly with the density $\rho$,
(ii) our calculations for $G_A$ can be approximated
globally by a dipole parametrization in the region $Q^2=0$--1.2 GeV$^2$;
(iii) above $Q^2=1.2$ GeV$^2$, we expect $G_A$
to have a falloff slower that a dipole parametrization
with $M_A \simeq 1.05$ GeV.

\subsubsection{Induced pseudoscalar form factor $G_P^\ast$}

The study of the falloff of $G_P^\ast$ must be taken with care
due the very rapidly changing function near $Q^2=0$,
discussed in the previous sections.
We make then a more qualitative analysis based
on a ratio to a value of $Q^2$
in a range where the falloffs of $G_P$ are more well defined.
We consider then the value $Q_p^2=0.012$ GeV$^2$
(still a small value, large enough to avoid the turning effect of 
the function $G_P^{\rm B}$).
Examples of calculations of $G_P^\ast(Q^2)/G_P^\ast(Q_p^2)$
are presented on the right side of Fig.~\ref{figGP-p30}
for $|\Delta I| =1$,
and on the right side of Figs.~\ref{figGP-p31} 
for $|\Delta S| =1$.
The calculations for all transitions are
  presented in Appendix~\ref{app-GP}
  (see Figs.~\ref{figGP-p30X} and \ref{figGP-p31X}).

The difference between the calculations
for different $\rho$ for the case $|\Delta I| =1$
(right side of Figs.~\ref{figGP-p30} and \ref{figGP-p30X}) 
are difficult to observe,
so we use a logarithmic scale.
The similarity of the calculations is a consequence of
the fact that the mass of the pion is unmodified by the medium and 
the factor $1/(m_\pi^2 + Q^2)$.
From the figures, one can conclude that, 
apart from the value of the function $G_P$ near $Q^2=0$,
the cases $\rho/\rho_0=0$, 0.5 and 1
are indistinguishable at low $Q^2$ ($Q^2 < 0.1$ GeV$^2$).
The ratio can then be interpreted as an universal function.
The corollary of this result
is that at low $Q^2$, we can determine the function
$G_P^\ast (Q^2)$ correcting the function in vacuum $G_P (Q^2)$ by
$G_P^\ast(Q_p^2)/G_P(Q_p^2)$.

An additional conclusion can be obtained when we
represent all ratios $G_P^\ast(Q^2)/G_P^\ast(Q_p^2)$ in one figure.
All lines are almost indistinguishable
below $Q^2=0.1$ GeV$^2$.
This result means that, in that range $G_P^\ast(Q^2)/G_P^\ast(Q_p^2)$
is a universal function (flavor independent)
that can be estimated using the results for the nucleon.

The results for  $|\Delta S| =1$
(right side of Figs.~\ref{figGP-p31} and \ref{figGP-p31X}),
now in a linear scale, show a different behavior from $|\Delta I| =1$ cases,
as a consequence of the factors $1/(m_K^{* 2} + Q^2)$.
There are two typical cases,
divided into light baryons ($\Lambda$ and $\Sigma$ decays) 
and into heavy baryons ($\Xi$ decays).
In the first case, for the 
$\Lambda \to p$ and $\Sigma^- \to n$/$\Sigma^0 \to p$ transitions, 
the ratios are almost the same near $Q^2= Q_p^2$ 
but differ in the range $Q^2 \simeq 0.02$--0.2 GeV$^2$,
till the lines converge again.
In that region the ratios are smaller for larger densities.
In the second case, for the 
$\Xi^- \to \Lambda$ and $\Xi^- \to \Sigma^0$ transitions
the ratios are the same below $Q^2=0.5$ GeV$^2$, and differ
after that point (larger ratio for larger densities).
Differently from the case $|\Delta I| =1$, 
the different transitions are characterized
by similar but different slopes
for $G_P^\ast(Q^2)/G_P(Q_p^2)$ at low $Q^2$.

The overall conclusion is that the functions
$G_P^\ast(Q^2)/G_P(Q_p^2)$ have a very sharp variation at low $Q^2$,
as a consequence of the significant effect of
the pole terms associated with the pion and the kaon.
In the case of $|\Delta I| =1$, we may expect that
the ratio $G_P^\ast(Q^2)/G_P(Q_p^2)$ is the same
for all transitions at \mbox{low $Q^2$}.

The description of the function $G_P^\ast$ can
be improved in the future, once stronger constrains
are included in the determination of the component $G_P^{\rm B}$,
and on the axial form factors, in general.

\subsection{Summary of the results in medium \label{sec-Summary1}}

We can summarize now the numerical results
for the weak interaction axial form factors, 
for the $|\Delta I| =1$ and $|\Delta S| =1$ transitions.

\subsubsection*{Axial-vector form factor $G_A^\ast$}

From the discussion in Sec.~\ref{secGA-star}, we conclude that
the axial-vector form factors are suppressed (quenched) in medium,
since $G_A^\ast/G_A < 1$ for the densities $\rho=0.5 \rho_0$ and $\rho_0$.
This suppression is a consequence of a reduction of
the meson cloud contribution, more significant at low $Q^2$,
and the bare contribution, more significant at large $Q^2$.

The suppression is stronger for light baryons
(like in the $n \to p$ and $\Lambda \to p$ transitions) than for heavy baryons
(like in the $\Xi^- \to \Xi^0$, $\Xi^0 \to \Sigma^+$
and $\Xi^- \to \Sigma^0$ transitions) 
and increases with the density.
At low $Q^2$, there is, in general, a stronger suppression
of $G_A$ due to the suppression of the meson cloud contribution.
In the case of light baryons, one can have
a substantial suppression for intermediate $Q^2$
(about 30\% for $n \to p$ and 25\% for
$\Lambda \to p$ for $Q^2=2$ GeV$^2$ and $\rho=\rho_0$).
In the case of the heavy baryons one has milder suppressions.
For the transitions $\Xi^- \to \Xi^0$,
$\Xi^0 \to \Sigma^+$ and  $\Xi^- \to \Sigma^0$
one can observe a suppression of 8\%--12\% at  $\rho=\rho_0$,
with a weak $Q^2$ dependence for the ratio $G_A^\ast/G_A$.
For the intermediate mass cases
(transitions $\Sigma^+ \to \Lambda$, $\Sigma^0 \to \Sigma^+$,
$\Sigma^- \to n$, $\Sigma^0 \to p$ and $\Xi^- \to \Lambda$)
we observe intermediate suppressions.

Except for a slightly stronger suppression
on the \mbox{$|\Delta I| =1$} transitions,
there are no significant differences in the
magnitudes of $G_A^\ast/G_A$  
between the $|\Delta I| =1$ and $|\Delta S| =1$ transitions.

\subsubsection*{Induced pseudoscalar form factor $G_P^\ast$}

We summarize now the results from Sec.~\ref{secGP-star}
for the induced pseudoscalar form factor in medium.
In general, one observes also a reduction of the ratio
$G_P^\ast/G_P$ in the nuclear medium.
In the present case, however, we need to distinguish between the
results for the $|\Delta I| =1$ and $|\Delta S| =1$ transitions.
The magnitude of the suppression increases with the density.

The difference between the $|\Delta I| =1$ and $|\Delta S| =1$ transitions 
is due to the dependence of the pole and meson cloud
terms on the mass of the pion ($|\Delta I| =1$)
or the mass of the kaon ($|\Delta S| =1$).
As a consequence, $G_P$ has larger magnitudes near $Q^2=0$
for the $|\Delta I| =1$ transitions than for the $|\Delta S| =1$ transitions 
(the results for $|\Delta I| =1$ are about 10 times
larger than the results for  $|\Delta S| =1$).
Another difference between the $|\Delta I| =1$ and $|\Delta S| =1$ transitions
is that $G_P$ displays a much slower falloff with $Q^2$
for $|\Delta S| =1$, due to the factor $1/(m_K^{* 2} + Q^2)$
on the pole and meson cloud terms
(the in-medium kaon mass is still larger than the pion mass).

In the case of $|\Delta I| =1$, the dominant term is the pole term.
In these conditions, one can expect that
$G_P^\ast/G_P \propto (M_{BB'}^\ast/M_{BB'})^2 (G_A^\ast/G_A)$. 
The consequence of this relation 
is that $G_P$ is suppressed in the nuclear medium 
and that the magnitude of the suppression
is determined by the reduction of the masses,
by the factor $ (M_{BB'}^\ast/M_{BB'})^2$
and by the reduction of the $G_A^\ast$ in medium (factor $G_A^\ast/G_A$).
The combination of the two effects is 
then a strong suppression for decays of light baryons
and a not so strong suppression for decays of heavy baryons
(smaller suppression factors).

As for the $|\Delta S| =1$ transitions,
one observes also a reduction of $G_P$ in medium  
but with different trends for low and intermediate/large $Q^2$.
At low $Q^2$, the magnitude of $G_P^\ast/G_P$
is smaller than the $|\Delta I| =1$ transitions,
because in this case the reduced mass of the kaon tends to increase
the magnitude of $G_P$ in-medium [$G_P^\ast/G_P \propto (m_K/m_{K}^\ast)^2$],
canceling partially the reduction due to the baryon masses and $G_A^\ast/G_A$.
For larger values of $Q^2$, however, the variation of $m_K^*$ became irrelevant
and we recover the relation $G_P^\ast/G_P \propto (M_{BB'}^\ast/M_{BB'})^2
(G_A^\ast/G_A)$
(valid for $|\Delta I| =1$).
As a consequence, there is a significant suppression
for light baryons and a very small suppression for heavy baryons
(with almost no dependence on $Q^2$ at large $Q^2$).

In both cases ($|\Delta I| =1$ and $|\Delta S| =1$),
the suppression on $G_P$ is stronger at low $Q^2$, due to the effect
of the meson cloud (stronger suppression of $G_A$ in medium).

Combining the cases $|\Delta I| =1$ and $|\Delta S| =1$
one can have at intermediate $Q^2$
($Q^2 \approx 1$ GeV$^2$) 
and $\rho=\rho_0$,
reductions of about 30--40\% for light baryons 
($n \to p$, $\Lambda \to p$, $\Sigma^- \to n$
and $\Sigma^0 \to p$)
and  about 5\%--10\% for heavy baryons
($\Xi^- \to \Xi^0$, $\Xi^- \to \Sigma^0$
and $\Xi^0 \to \Sigma^+$).
For intermediate mass baryons
($\Sigma^+ \to \Lambda$, $\Sigma^0 \to \Sigma^+$,
$\Sigma^- \to n$, $\Sigma^0 \to p$ and $\Xi^- \to \Lambda$)
one has intermediate reductions (15\%--20\%).

\subsection{Discussion \label{sec-discussion}}    

The results of the previous sections for $G_A$ and $G_P$,
in vacuum and in the medium, are a natural consequence
of the formalism proposed and discussed
in Secs.~\ref{secVacuum} and \ref{secMedium}.
As a consequence, the properties of the functions $G_A$ and $G_P$
are the by-product of the assumptions and methods used. 
We discuss here the virtues and the
limitations of the present calculations.

Some of the results for $G_A$ and $G_P$ mentioned in the previous sections
are the outcome of the methodology used to determine the 
free parameters of the model to the lattice QCD data
for the nucleon~\cite{AxialFF,Alexandrou11a}.
The calibration based on the lattice QCD data
has been limited by the lack of data between $Q^2=0$ and $Q^2 =0.15$ GeV$^2$.
This lack of data has consequences on the determination
of the $n_P$ and on the quark form factor $g_P^q$, and
may affect the accuracy of the calibration of the functions  
$G_A$ and $G_P$ at low $Q^2$.

We first comment on the results for $G_A$.
As mentioned already, the weak axial-vector form factor
is very smooth near $Q^2=0$.
This property was already present in the model for the vacuum~\cite{AxialFF},
but was not so clearly observed due to the competition between
the bare contribution and the meson cloud contribution
with a sharp shape  $\propto 1/(1 + Q^2/M_A^2)^4$.
The smooth shape of the bare contribution can be seen
in the bare contribution in Fig.~\ref{figGAGP-Nucleon} for the nucleon,
and it is a consequence of $P$-wave contributions.
Since the $P$-state mixture was determined by a fit to 
lattice and physical data in a range $Q^2=0$--3 GeV$^2$
(equal weight for all $Q^2$ points),
no special constraint was included in the fit to the data near $Q^2=0$.
This is the main reason why the square axial radius deviates
from the estimate based on a dipole function.
In medium, the function $G_A$ has become softer when $|n_P|$ increases.
The present results for $G_A$ are then mainly a consequence
of the determination of the parameters of the model based on the
lattice QCD simulations, where the low-$Q^2$ region is not well represented,
due to the gap of 0.3 or 0.4 GeV$^2$
between the lattice QCD data points~\cite{Alexandrou11a}.
The properties of the function $G_A$ for the nucleon are extended
to the octet baryon case.

Also related to the previous topic are the results for $G_P$ at low $Q^2$.
Contrarily to the functions $G_A$, there are no lattice QCD data
for $Q^2=0$ (as there are no lattice QCD data for the Pauli form factor $F_2$).
Since the function $g^q_P(Q^2)$, defined in Eq.~(\ref{eqQuarkGP}),
contributes only to the bare contribution in the induced pseudoscalar
form factor $G_P$ [see Eq.~(\ref{eqGPoct})],
the free parameters of $g^q_P(Q^2)$ are determined
exclusively by the fit to the $G_P$ lattice QCD data.
In these conditions, the calibration of the function $g_P^q$
is the result of the data above the minimum value for $Q^2$
($Q^2 > 0.15$ GeV$^2$)
which may significantly affect the result for $G_P$ near $Q^2=0$, 
since that is the point where the global function $G_P$
is expected to have its maximum.
More recent data with a wide range in $Q^2$,
or independent estimates of $G_P(0)$ may help
to improve the calibration of $G_P^{\rm B}$ for $|\Delta I| =1$  and $|\Delta S| =1$.
In the alternative, a fitting procedure with some theoretical
constraints to the function $G_P$ at low $Q^2$ may also
be used in future studies.

To finish the discussion about the present calculations
we comment on possible improvements in the model.
As mentioned already, the formalism was applied first
to the calculation of the electromagnetic form factors
in the nuclear medium using the simplest form for
the octet baryon wave functions,
restricted to an $S$-state component~\cite{Octet2,Octet3}.
In a previous work~\cite{AxialFF}, we concluded, however,
that for the study of the axial form factors
in addition to the $S$-state, it was also necessary 
a $P$-state component.
The contributions of the meson cloud effects,
the octet baryon wave functions,
assumed to be dominated by the pion cloud effects,
were determined phenomenologically
by the electromagnetic form factor data.
In the study of the axial structure, we have considered
the formalism derived from the electromagnetic structure,
with different proportion of the pion cloud contribution
(different pion-nucleon self-energy parameter $b_1$).
In the future, we should consider a global
parametrization, where the electromagnetic
and axial structure form factors of the octet baryons are calculated
using the same wave functions, and the same proportion
of the meson cloud in the normalization of the baryon states.

The parametrization of the meson cloud contributions
at the level of the electromagnetic structure
can also be improved with the inclusion of terms associated
with the kaon cloud for the $\Lambda$, $\Sigma$ and $\Xi$ systems.
The inclusion of the new terms will be sufficient to improve
the description of the $\Sigma$ and $\Xi$ weak decay data, 
because it reduces the value of 
$\sqrt{Z_B}$ and, consequently, the magnitude of the $G_A$ 
(see Sec.~\ref{sec-Vacuum}).

In the future, we may also consider a less phenomenological
and more physical motivated description of the meson cloud contribution 
for the axial form factors, instead
of the simplified SU(3) baryon-meson interaction
(functions $F'$ and $D'$).
That can be done considering a microscopic connection between
the covariant spectator quark model and the bag model
relating the electromagnetic-baryon
and meson-baryon couplings of the two frameworks
as in Refs.~\cite{DecupletDecays,OctetDecupletD2}.

The methodology discussed in the present work combined
with the models developed for the decuplet baryon 
and the octet baryon to decuplet
baryon transitions~\cite{Omega,OctetDecuplet1,DecupletDecays,OctetDecupletD1,OctetDecupletD2}
can be used to extend the model calculations 
to the study of the electromagnetic and axial structure
of decuplet baryons and the octet baryon to decuplet baryon transitions
in nuclear matter.

%%  VSPACE

\section{Numerical results for neutrino-nucleus cross sections 
\label{secResultsCS}}

In the present section, we exemplify how the
weak axial form factors calculated in the
previous section, can be used to calculate
neutrino-nucleus and antineutrino-nucleus
cross sections considering the interaction
of a neutrino or antineutrino with nucleons immersed
in a nuclear matter with density $\rho$.
In addition to the form factor $G_A$ we need
also to know the electric ($G_E$) and magnetic ($G_M$) form factors in medium.
In that case we consider the model from Ref.~\cite{Octet2}
also derived within the covariant spectator
quark model formalism.

The interactions with neutrinos/antineutrinos
can be divided in neutral current (NC)
and charged current (CC) reactions 
where leptons ($e,\mu$) can be found in the final state.
Here we use $M$ to represent the nucleon mass in vacuum or in medium.
For convenience, we analyze the reaction
at the Lab frame, where the nucleon (momentum $p$)
is at rest and the neutrino/antineutrino (momentum $k$)
has an energy $E_\nu$, and the transfer momentum to the nucleon is $q$.
In these conditions, the transition cross sections
can be expressed in terms of the invariants $Q^2$, 
$y= \frac{p\cdot q}{p \cdot k} = \frac{Q^2}{2 M E_\nu}$,
the energy $E_\nu$, and the nucleon
electromagnetic and axial form factors, as described below.

The single-differential cross section for NC reactions
($\nu N \to \nu N$ and $\bar \nu N \to \bar \nu N$)
takes the form~\cite{Cheoun13b,Alberico02a,Meucci04a}
\ba
\left( \frac{d \sigma}{d Q^2} \right)_{\nu ({\bar \nu})}^{\rm NC} &= &
\frac{G_F^2}{2\pi} \left[ \frac{1}{2} y^2 \; (G_M^{\rm NC})^2 \right. \nonumber \\
  & &  %\left.
  + \left( 1 -y - \frac{M}{2 E_\nu} y \right)
  \frac{(G_E^{\rm NC})^2 + \frac{E_\nu}{2M} y  \; (G_M^{\rm NC})^2}{1 + \frac{E_\nu}{2M} y }
  \nonumber \\
  && +  \left( \frac{1}{2} y^2 + 1 -y + \frac{M}{2 E_\nu} y\right)\;  (G_A^{\rm NC})^2
  \nonumber \\
  & & \left. \mp 2y \left(1 - \frac{1}{2}y \right) G_M^{\rm NC} G_A^{\rm NC} \right],
\label{eqSigmaNC-bar-nu}
\ea
where $G_F = 1.664 \times 10^{-5}$ GeV$^{-2}$ is the Fermi constant, 
and $G_\ell^{\rm NC}$ ($\ell = E,M$) and $G_A^{\rm NC}$
are the NC form factors, as defined next.
The upper sign in the last term corresponds to
the reactions with neutrinos and the lower sign the reaction 
with antineutrinos.

The NC form factors take the form  ($\ell = E,M$)
\ba
G_\ell^{\rm NC} =
\left\{
\begin{array}{cc} \frac{t}{2} G_\ell^p - \frac{1}{2} G_\ell^n & p \; \mbox{target} \cr
                  \frac{t}{2} G_\ell^n - \frac{1}{2} G_\ell^p & n \; \mbox{target}  \cr 
\end{array}
\right. ,
\label{eqGellNC}
\ea
with  $t= 1 - 4 \sin^2 \theta_{\rm w} \simeq 0.076$,
($\theta_{\rm w}$ is the weak mixing or Weinberg angle) and
\ba
G_A^{\rm NC}=
\left\{
\begin{array}{cc} - \frac{1}{2} G_A^{np} & p \; \mbox{target} \cr
                  \frac{1}{2} G_A^{np} & n \; \mbox{target}  \cr 
\end{array}
\right. .
\label{eqGANC}
\ea
The function $G_A^{np}$ is the nucleon axial-vector form factor
(associated with the $n \to p$ decay) discussed in the previous sections.

The single-differential cross sections for
the CC reactions ($\nu_e n \to e^- p$ and 
$\bar \nu_e p \to e^+ n$)  
can be also written in a similar form
(assuming that the electron mass can be neglected)~\cite{Cheoun13a,Alberico02a,Meucci04a}
\ba
\left( \frac{d \sigma}{d Q^2} \right)_{\nu ({\bar \nu})}^{\rm CC} &= &
\frac{G_F^2}{2\pi} \left[ \frac{1}{2} y^2 \; (G_M^{\rm CC})^2 \right. \nonumber \\
  & &  %\left.
  + \left( 1 -y - \frac{M}{2 E_\nu} y \right)
  \frac{(G_E^{\rm CC})^2 +\frac{E_\nu}{2M} y \; (G_M^{\rm CC})^2}{1 + \frac{E_\nu}{2M} y}
  \nonumber \\
  && +  \left( \frac{1}{2} y^2 + 1 -y + \frac{M}{2 E_\nu} y \right)\;  (G_A^{\rm CC})^2
  \nonumber \\
  & & \left. \mp 2y \left(1 - \frac{1}{2}y \right) G_M^{\rm CC} G_A^{\rm CC} \right],
\label{eqSigmaCC-bar-nu}
\ea
where ($\ell = E,M$)
\ba
G_\ell^{\rm CC} =  G_\ell^p - G_\ell^n,
\label{eqGellCC}
\ea
and
\ba
G_A^{\rm CC}= - G_A^{np}.
\label{eqGACC}
\ea
Again, the signs in the last term, respectively, correspond to
the reactions with neutrinos (upper sign) or antineutrinos (lower sign).

The previous relations use the magnetic form factors $G_M$
in natural units: $G_M= F_1 + F_2$, where $F_1$, $F_2$
are the Dirac and Pauli form factors respectively.

In this analysis, we neglect for simplicity the
strange quark contributions for the electromagnetic form factors ($G_E^s$, $G_M^s$)
and axial form factors ($G_A^s$)~\cite{Cheoun13a,Cheoun13b,Meucci04a}.
For a discussion about the effect of the strange quark
see Refs.~\cite{Alberico02a,Pate24a}.

The previous expressions are valid for the vacuum using $M$
as the nucleon mass in free space,
and for a nuclear medium replacing the nucleon mass by $M_N^\ast$
and the form factors ($G_\ell$ and $G_A$) by the in-medium
form factors ($G_\ell^\ast$ and
$G_A^\ast$).
In the nuclear medium the value of $y$ is also corrected to
\ba
y = \frac{Q^2}{2 M^\prime E_\nu},
\label{eqYmod}
\ea
where $M^\prime = M_N^\ast + V$ and $V$ is a vector potential term
associated with the nucleon-medium interaction.
Numerically one has $V= 0.0627$ GeV for $\rho=0.5 \rho_0$
and $V=0.1253$ GeV for $\rho= \rho_0$.
The last correction is a consequence of the modification of the
nucleon energy at rest in medium, where the nucleon momentum
takes the form $p=(M^\prime, {\bf 0})$~\cite{QMCReview}.
The transfer momentum $q$ is then modified in medium\footnote{The transfer momentum
  takes the form $q=(\omega,{\bf q})$,
  where $\omega= \frac{Q^2}{2 M^\prime}$ and
  $|{\bf q}|^2= Q^2 \left(1 + \frac{Q^2}{4 M^{\prime 2}} \right)$.}
leading to the result (\ref{eqYmod}).

The expressions for NC reactions can be generalized from
electrons for muons ($\nu_e$ to $\nu_\mu$).
The generalization of the CC reactions from electrons to
muons is only approximated due to the (large) mass of the muon.

The transitions mentioned above can be divided
into the following categories:
\begin{itemize}
\item
  NC transitions (proton target): \\
    $\nu p \to \nu p$ and
    $\bar \nu p \to \bar \nu p$,
    \item
      NC transitions (neutron target): \\
      $\nu n \to \nu n$ and $\bar \nu n \to \bar \nu n$, 
  \item
    CC transitions: \\
     $\nu n \to e^- p$ and $\bar \nu p \to e^+ n$.
\end{itemize}  

The magnitudes of the cross sections for
these three cases are a consequence of the magnitude
of the form factors $G_\ell$ and $G_A$.
We start by looking for the form factors
for the three kinds of transitions.

\begin{figure*}[t]
  \centerline{ \mbox{
\includegraphics[width=2.2in]{GE-NCpR} \hspace{.15cm}
\includegraphics[width=2.2in]{GM-NCpR} \hspace{.15cm}
\includegraphics[width=2.2in]{GA-NCp}}}
  \centerline{ \vspace{.5cm}}
    \centerline{ \mbox{
\includegraphics[width=2.2in]{GE-NCnR} \hspace{.15cm}
\includegraphics[width=2.2in]{GM-NCnR} \hspace{.15cm}
\includegraphics[width=2.2in]{GA-NCn}}}
     \centerline{ \vspace{.5cm}}
      \centerline{ \mbox{
\includegraphics[width=2.2in]{GE-CC} \hspace{.15cm}
\includegraphics[width=2.2in]{GM-CC} \hspace{.15cm}
\includegraphics[width=2.2in]{GA-CC}}}
\caption{\footnotesize{Electric, magnetic and axial form factor
    associated with different reactions.
    {\bf At the top:} NC reactions with proton targets.{\bf At the center:}
    NC reactions with neutron targets.
{\bf At the bottom:} CC reactions.}}
\label{fig-FFs}    
\end{figure*}

\subsection{Form factors for NC and CC reactions  \label{secGEGMGA}}

The form factors for NC transitions are
defined by Eqs.~(\ref{eqGellNC}) and (\ref{eqGANC})
for proton and neutron targets,
and are presented at the top (proton) 
and central (neutron) panels in Fig.~\ref{fig-FFs}, respectively.
The form factors for the CC transitions are defined by
Eqs.~(\ref{eqGellCC}) and (\ref{eqGACC})
and are presented at the bottom panel in Fig.~\ref{fig-FFs}.

The results from Fig.~\ref{fig-FFs} exhibit a reduction of the 
magnitude in the form factors
(absolute values) in the nuclear medium.
The reduction increases 
with the density $\rho$ (the form factors are quenched in medium).
These results are a consequence of the combination of proton and neutron form factors.
As discussed in Refs.~\cite{Octet2,Octet3} proton and neutron form factors
are modified differently in the nuclear medium.
The exceptions to the suppression are the results of the electric
and the magnetic form factors at $Q^2=0$.
The suppression of $G_A$ in medium was already discussed in Sec.~\ref{secMedium}.

At $Q^2=0$ the electric form factor is unmodified by the medium
because the electric charge does not change in medium.
As for the magnetic form factor at $Q^2=0$, the invariance of $G_M^{\rm NC}(0)$
is only approximated, and it is a consequence of the general 
dominance of the valence quark contributions in the nuclear medium~\cite{Octet2}.
[The quark anomalous magnetic moments in natural units are not modified in medium.]

The medium effect on the magnetic form factors has a different
trend when expressed in terms of the nuclear magneton in vacuum
[units $e/(2M_N)$], instead of the natural units, as discussed
in Ref.~\cite{Octet2}.
In the first case the nucleon magnetic form factors in medium are corrected
by the factor $\frac{M_N}{M_N^\ast}$~\cite{Octet2}.
The consequence of this conversion is that
since  $G_M^{\rm NC}(0)$ in natural units is almost independent
of the medium, $G_M^{\rm NC}(0)$ is enhanced when expressed in units $e/(2M_N)$.

For the cases under discussion, the functions $G_M^{\rm NC}(Q^2)$
and $G_M^{\rm CC}(Q^2)$ from  Fig.~\ref{fig-FFs}
(combinations of proton  and neutron form factors),
we can conclude that the amplification due to the factor
$\frac{M_N}{M_N^\ast}$ dominates over the suppression of $G_M^{\rm NC}(Q^2)$
in natural units.
In conclusion, $G_M(Q^2)$ is enhanced in medium at low $Q^2$, 
when we use units $e/(2M_N)$.
When expressed in natural units, $G_M^{\rm NC}(Q^2)$ and $G_M^{\rm NC}(Q^2)$
are suppressed in medium.

The results from Fig.~\ref{fig-FFs} show that the function $G_E^{\rm NC}$
is very small for proton targets, as a consequence of
the small magnitude of $1- 4 \sin^2 \theta_{\rm w}$.
For the discussion, we also notice that
the magnitude of the form factors
$G_E$ and $G_A$ for CC reactions are about
2 times the magnitudes for NC reactions with neutron targets.
This effect has an impact on the magnitude of the
neutrino-nucleon cross sections, as discussed later.

It is worth mentioning that the single-differential cross sections
in terms of $Q^2$ and the neutrino energy $E_\nu$ are not directly measured.
The comparison between the theoretical expressions
and experimental data for $\frac{d \sigma}{d Q^2}$
in terms of $Q^2$ requires the knowledge
of the neutrino/antineutrino energy spectrum,
since measurements are made for defined ranges of
neutrino/antineutrino energies~\cite{Alberico02a,Ahrens87a}.
In most cases, measurements include the scattering of electron
and muon neutrinos~\cite{Alberico02a,Ahrens87a,Super-Kamiokande99,NOvA23}.

%\newpage

\subsection{Kinematics}

In the calculation of the single-differential cross sections
one needs to take into account the kinematic relations between
the neutrino energy $E_\nu$ and $Q^2$.
As before we consider the Lab frame
(initial nucleon at rest).

The squared momentum transfer $Q^2$ for the neutrino-nucleon
scattering with a neutrino energy $E_\nu$
and an angle of $\theta_{l}$ between the outgoing
and incoming neutrino is~\cite{Hoferichter20a}
\ba
Q^2= \frac{2 E_\nu^2 (1 - \cos \theta_{l})}{1 +
\frac{E_\nu}{M} (1 - \cos \theta_{l}) },
\ea
neglecting the neutrino mass.
The same analysis is valid for antineutrino reactions.
From the previous relation we can conclude that
the range of values of $Q^2$ according to the
possible values of $u = \cos \theta_{l}$: $-1 \le u \le 1$
is in the interval
\ba
0 \le Q^2 \le \frac{4 M E_\nu^2}{ M + 2 E_\nu}.
\label{eqQ2-limits}
\ea
The same conclusion is obtained
when the analysis is made at the center of mass of
the neutrino-nucleon system~\cite{Chen24a}.

The upper limit ($Q^2_{\rm max}$) defines the maximum value of $Q^2$ that
can be obtained for the neutrino energy $E_\nu$.
One can then conclude that for $E_\nu=5$ MeV
one has $Q^2_{\rm max} \simeq 10^{-4}$ GeV$^2$,
for $E_\nu=20$ MeV, one has $Q^2_{\rm max} \simeq 1.5 \times 10^{-3}$ GeV$^2$,
and for $E_\nu=100$ MeV, one has $Q^2_{\rm max} \simeq 0.035$ GeV$^2$.
Higher values  can be obtained for $Q^2_{\rm max}$ 
when we consider the scattering of
muon neutrinos/antineutrinos ($\nu_\mu/\bar \nu_\mu$)
instead of electron neutrinos/antineutrinos ($\nu_e/\bar \nu_e$)~\cite{KDAR}.
In these estimates we use the nucleon mass for the
density $\rho=0.5 \rho_0$ as an example.

The relation $E_\nu$--$Q^2_{\rm max}$ explains why
the last term of the NC and CC single-differential cross sections
(\ref{eqSigmaNC-bar-nu}) and (\ref{eqSigmaCC-bar-nu})
does not change sign when the sign
of $G_M^{\rm NC} G_A^{\rm NC}$ and $G_M^{\rm CC} G_A^{\rm CC}$ is preserved.
The conditions (\ref{eqQ2-limits}) are equivalent to $0 \le y < 1$.
[In that range, the factor $1 - \frac{1}{2} y$ is always positive,
or zero (when $Q^2=0)$.]

The expressions mentioned above can be used also
to determine the value of the neutrino energy $E_\nu$
necessary to obtain a given value $Q^2_{\rm max}$:
\ba
E_\nu = M \left[ \sqrt{\tau_m + \tau_m^2} + \tau_m \right],
\ea
where $\tau_m = \frac{Q^2_{\rm max}}{4 M^2}$.

We conclude then that we need $E_\nu= 100$ MeV 
to reach $Q^2_{\rm max}= 0.035$ GeV$^2$, and $E_\nu =0.88 $ GeV
to reach $Q^2_{\rm max}= 1$ GeV$^2$.
Again, we use the values of the nucleon mass for  $\rho=0.5 \rho_0$.

%\newpage

\subsection{Cross sections for NC reactions}

The NC transitions for proton and neutron targets
are presented in Figs.~\ref{fig-sigmaNCp} and \ref{fig-sigmaNCn},
respectively, for the densities
$\rho=0$, $0.5 \rho_0$ and $\rho_0$.
In the upper panels we represent
the $\nu N \to \nu N$ transitions
and in the central panels the $\bar \nu N \to \bar \nu N$ transitions.
The vertical lines represent the upper limit of $Q^2$
associated with the neutrino energies $E_\nu=5$, 20 and 100 MeV.
The results for the different energies can
be inferred from the upper limit for $Q^2$.

The first conclusion from the observations of the figures
is that the single-differential cross sections are reduced
in medium (quenched effect) and that the suppression
is more significant for higher densities.
These results are a consequence of the suppression
of the form factors $G_\ell^{\rm NC}$ and  $G_A^{\rm NC}$ in medium,
discussed earlier.
The difference of magnitudes between proton and
neutron reactions is a consequence of the
magnitude of the form factors [see Fig.~\ref{fig-FFs}].

The second conclusion is that there is a dominance
of the cross sections for neutrinos (upper panel)
over the cross sections for antineutrinos (center panel).
The dominance of the reactions with neutrinos over
the antineutrinos is a general property of
the $\nu N$ and $\bar \nu N$ reactions.
The result is a consequence of the
last term in Eq.~(\ref{eqSigmaNC-bar-nu}),
where neutrino contributions (sign $-$) are
positive, because $G_M^{\rm NC} G_A^{\rm NC} < 0$,
while the contributions for the antineutrino (sign $+$)
are negative.

For a better visualization of this effect, we compare
at the bottom panel in Figs.~\ref{fig-sigmaNCp} and \ref{fig-sigmaNCn}
the results for $\nu N \to \nu N$ 
the $\bar \nu N \to \bar \nu N$ for a fixed density
($\rho = 0.5 \rho_0$).
Similar results are obtained for $\rho = 0$ and $\rho= \rho_0$.

In the limit $Q^2=0$ the neutrino and antineutrino
cross sections have the same value because
the last terms of Eqs.~(\ref{eqSigmaNC-bar-nu}) and (\ref{eqSigmaCC-bar-nu})
vanishes.
The magnitudes of the cross sections are larger for $E_\nu =20$ MeV.
The variation of the antineutrino cross sections with $Q^2$ for larger values of $E_\nu$:
almost constant for proton targets and a soft reduction with
neutron targets is a consequence of the dependence of
$G_M^{\rm NC}$ with $Q^2$.
The variation of the antineutrino cross sections with $Q^2$:
almost constant for proton targets and the reduction with
neutron targets is a consequence of the dependence of
$G_M^{\rm NC}$ with $Q^2$.
The product $G_M^{\rm NC} G_A^{\rm NC}$ is negative for protons and neutrons
but the magnitude is larger for the case of the neutrons, 
because $|G_M^{\rm NC}|$ is larger for neutron targets
($|G_A^{\rm NC}|$ is the same for both cases).
As a consequence, the antineutrino single-differential cross section is reduced
for the neutron target when $Q^2$ increases.

%%%    Figures 12, 13

%%    Sizes:  3.05, 3.00   too big

\begin{figure}[t]
\vspace{.45cm}
\centerline{
\mbox{
\includegraphics[width=2.95in]{dsig-nu-p-v2}
}}
\centerline{\vspace{.5cm} }
\centerline{
\mbox{
\includegraphics[width=2.95in]{dsig-nubar-p-v2}
}}
\centerline{\vspace{.5cm} }
\centerline{
\mbox{
\includegraphics[width=2.95in]{dsigNC-p-compare}
}}
\caption{\footnotesize{Single-differential cross sections
for NC reactions with proton targets
for the densities $\rho=0$, $0.5 \rho_0$, $\rho_0$.
Neutrino energies: $E_\nu=5$, 20 and 100 MeV.
{\bf At the top:} Neutrino-proton scattering $\nu p \to \nu p$.
{\bf At the center:}
Antineutrino-proton scattering $\bar \nu p \to \bar \nu p$.
{\bf At the bottom:} Comparison between
$\nu p \to \nu p$ and $\bar \nu p \to \bar \nu p$ transitions.
$\rho= 0.5 \rho_0$.
}}
\label{fig-sigmaNCp}
\end{figure}

\begin{figure}[t]
\vspace{.45cm}
\centerline{
\mbox{
\includegraphics[width=2.95in]{dsig-nu-n-v2}
}}
\centerline{\vspace{.5cm} }
\centerline{
\mbox{
\includegraphics[width=2.95in]{dsig-nubar-n-v2}
}}
\centerline{\vspace{.5cm} }
\centerline{
\mbox{
\includegraphics[width=2.95in]{dsigNC-n-compare}
}}
\caption{\footnotesize{
Single-differential cross sections
for NC reactions with neutron targets
for the densities $\rho=0$, $0.5 \rho_0$, $\rho_0$.
Neutrino energies: $E_\nu=5$, 20 and 100 MeV.
{\bf At the top:} Neutrino-neutron scattering $\nu n \to \nu n$.
{\bf At the center:}
Antineutrino-proton scattering $\bar \nu n \to \bar \nu n$.
{\bf At the bottom:} Comparison between
$\nu n \to \nu n$ and $\bar \nu n \to \bar \nu n$ transitions.
$\rho= 0.5 \rho_0$.
}}
\label{fig-sigmaNCn}
\end{figure}

%\newpage

\subsection{Cross sections for CC reactions}

The single-differential cross sections
for the CC reactions $\nu n \to e^- p$ and $\bar \nu p \to e^+ n$ 
are presented in Fig.~\ref{fig-sigmaCC}
for the densities
$\rho=0$, $0.5 \rho_0$ and $\rho_0$.
We consider the range of $Q^2$ associated
to the neutrino energies $E_\nu =5$, 20 and 100 MeV.
As before, the results for the different energies
can be inferred from the upper limit lines for $Q^2$.

As for the NC transitions, we conclude
that the  cross sections are reduced in medium (quenched effect).
Also for NC transitions one can observe the dominance of the neutrino cross sections
over the antineutrino cross sections.

In comparison with the NC transitions we can notice
that the scale of the single-differential cross sections
increases by a factor of about 4.
This result is a consequence of the
relations between the form factors $G_E$ and $G_A$
for NC reactions with neutron targets and CC reactions,
discussed previously.
The form factors for CC reactions are about 2 times larger.
The effect of $G_M$ in the cross sections
is reduced by the multiplication with~$y$.

The relation between the $\nu n \to e^- p$ and $\bar \nu p \to e^+ n$
cross sections confirm the role of the neutrino/antineutrino reactions
already discussed for $\nu N \to \nu N$ reactions 
(to be compared with  $\nu n \to e^- p$)
and $\bar \nu N \to \bar \nu N$ reactions
(to be compared with $\bar \nu p \to e^+ n$).

%%%    Figure 14

\begin{figure}[t]
%\vspace{.45cm}
\centerline{
\mbox{
\includegraphics[width=2.95in]{dsig-e-p-v2}
}}
\centerline{\vspace{.5cm} }
\centerline{
\mbox{
\includegraphics[width=2.95in]{dsig-e+n-v2}
}}
\centerline{\vspace{.5cm} }
\centerline{
\mbox{
\includegraphics[width=2.95in]{dsigCC-compare}
}}
\caption{\footnotesize{Single-differential cross sections
for CC reactions
for the densities $\rho=0$, $0.5 \rho_0$, $\rho_0$.
Neutrino energies: $E_\nu=5$, 20 and 100 MeV.
{\bf At the top:} $\nu n \to e^- p$ scattering.
{\bf At the center:}
$\bar \nu p \to e^+ n$ scattering.
{\bf At the bottom:} Comparison between
$\nu n \to e^- p$ and $\bar \nu p \to e^+ n$
single-differential cross sections.
$\rho= 0.5 \rho_0$.
}}
\label{fig-sigmaCC}
\end{figure}

%% HVSPACE

\newpage

\subsection{Summary}

In the present section, we exemplify how to
use the calculations of the electromagnetic and axial-vector
form factors in medium to calculate the single-differential cross sections
of NC and CC reactions in medium.
In general the cross sections are reduced in medium.
The magnitude of the calculations depends on the
incident particle (neutrino or antineutrino).
The calculation of the $\frac{d \sigma}{d Q^2}$
in terms of $Q^2$ requires the knowledge of
the neutrino/antineutrino energy $E_\nu$ distribution.

The present calculations are consistent with
the dominance of the neutrino cross sections 
over the antineutrino single-differential cross sections.
The effect was observed on the BNL experiments~\cite{Alberico02a,Ahrens87a}, 
where the averaged neutrino over the antineutrino
single-differential cross sections for muon neutrinos
($\nu_\mu p \to \nu_\mu p$ vs $\bar \nu_\mu p \to \bar \nu_\mu p$)
are determined.
The main difference to the present calculations is 
on the lepton family (muon neutrinos instead of electron neutrinos) and
the energy range of neutrinos
($E_\nu \simeq 1$ GeV, instead of $E_\nu \simeq 0.1$ GeV).

%\newpage

\section{Outlook and conclusions \label{secConclusions}}

The literature about the electromagnetic and axial structure
of baryons in the nuclear medium is scarce, 
except for the case of the nucleon.
This work is one more relevant contribution to the field.
We present a systematic study of the axial
form factors of the octet baryons in terms of the square transfer momentum ($q^2=-Q^2$)
and the density of the nuclear medium. 
We extend here previous studies in the free space for
octet baryons immersed in a nuclear medium.
The calculations presented can be used in the future studies
of interactions between the octet baryons with neutrinos or
antineutrinos in the nuclear medium, at low, intermediate and large $Q^2$.
We expect the present study to stimulate future works
on the subject and promote a debate between the different formalisms
used in the studies of the reactions in the nuclear medium.

Our calculations are based on the covariant spectator quark model,
which takes into account
the valence quark effects and the meson cloud dressing effects
on the electroweak interactions with baryon systems.
The model has been used successfully in 
studies of the electromagnetic and axial structure
of several baryon systems including resonances of the nucleon
in free space.
The extension of the model to the nuclear medium takes
into account the modifications of the properties of mesons and baryons
(masses and coupling constants) in the medium
determined by the quark-meson coupling model.

The valence quark component of the model uses
a vector meson dominance parametrization of the quark current
and the SU(6) flavor-spin symmetry to determine
the radial wave functions of the baryon systems. 
The free parameters of the model were determined
in the previous studies of the octet baryon systems
in lattice QCD and in the free space.
The meson cloud dressing is estimated phenomenologically
using the SU(3) baryon-meson model
and the physical information of the octet baryon axial couplings.

We use the formalism to calculate the weak axial-vector $G_A$
and induced pseudoscalar $G_P$ form factor associated
with the $B' \to B$ transitions between the octet baryon members,
for different nuclear densities, including the vacuum
($\rho=0$), and the normal nuclear matter density
($\rho=\rho_0 \simeq  0.15$ fm$^{-3}$).
The formalism can be used for other values of $\rho$ including $\rho > \rho_0$.

Our calculations in the nuclear medium suggest that
the weak axial-vector form factor $G_A$ is suppressed in the nuclear medium.
The suppression is stronger for light baryons and high densities.
The effect increases, in general, with $Q^2$.
In the case of the $n \to p$ transition about 30\%
for the nuclear density for $Q^2=2$ GeV$^2$.

As for the induced pseudoscalar form factor $G_P$,
because it is an observable that is hard to measure even in the vacuum,
we have made more qualitative estimates.
Our calculations suggest that $G_P$ is significantly suppressed in medium.
In the case of $|\Delta I| =1$ transitions (only $u$ and $d$ quarks are involved)
the suppression is more significant for light baryons
and for transitions with significant suppression in $G_A$.
In the case of  $|\Delta S| =1$ transitions (a strange quark
transition is involved),
one can have significant suppression at low $Q^2$ in general.
At large $Q^2$ the suppression can be relevant for light baryons
and very mild for heavy baryons (especially for $\Xi$).

To exemplify how the present calculations
can be used we calculate the single-differential cross sections
of the neutrino and antineutrino interactions with a
nucleon immersed in symmetric nuclear matter
with densities $\rho=0$, $0.5\rho_0$ and $\rho_0$.
In the future, we will consider the application
of the formalism for interactions
of neutrinos or antineutrinos with specific nuclei.

The present formalism can be extended in
the future to the study of the electromagnetic
and axial structure of the decuplet baryons
and for the octet baryon to decuplet baryon transitions.

%\vspace{.5cm}

\begin{acknowledgments}
  %\vspace{-.4cm}
G.~R.~thanks H.~Parada for useful discussions.
G.~R.~was supported by the Basic Science Research Program 
funded by the Republic of Korea Ministry of Education 
(Grant No.~NRF--2021R1A6A1A03043957).
K.T.~was supported by Conselho Nacional de Desenvolvimento 
Cient\'{i}fico e Tecnol\'ogico (CNPq, Brazil), Processes No.~313063/2018-4,
 No.~426150/2018-0 and No.~3014199/2022-2,
 and FAPESP Processes No.~2019/00763-0
and No.~2023/07313-6, and his work was also part of the projects,
Instituto Nacional de Ci\^{e}ncia e 
Tecnologia - Nuclear Physics and Applications 
(INCT-FNA), Brazil, Process No.~464898/2014-5.
The work of  M.~-K.~C.~is supported by the National
Research Foundation of Korea (Grants Nos.~NRF--2021R1A6A1A03043957 and NRF--2020R1A2C3006177).
\end{acknowledgments}

%\newpage

%%  XVSPACE

%\vspace{.5cm}

\appendix

\section{Gell-Mann matrices}
\label{appGM-matrices}

To complement the discussion of the main text 
we present here the explicit form of the Gell-Mann
matrices~\cite{Gaillard84}
{\small\ba
\hspace{-.5cm}
&&
\lambda_1= 
\left(
\begin{array}{ccc} 0 & 1 & 0 \cr 
                   1 & 0 & 0 \cr
                   0 & 0 & 0 \cr
\end{array}
\right),
\hspace{.1cm}
\lambda_2= 
\left(
\begin{array}{ccc} 0 & -i & 0 \cr 
                   i & 0 & 0 \cr
                   0 & 0 & 0 \cr
\end{array}
\right),
\hspace{.1cm}
\lambda_3= 
\left(
\begin{array}{ccc} 1 & 0 & 0 \cr 
                   0 & -1 & 0 \cr
                   0 & 0 & 0 \cr
\end{array}
\right), \nonumber \\
\hspace{-.5cm}
&& 
\lambda_4= 
\left(
\begin{array}{ccc} 0 & 0 & 1 \cr 
                   0 & 0 & 0 \cr
                   1 & 0 & 0 \cr
\end{array}
\right),
\hspace{.1cm}
\lambda_5= 
\left(
\begin{array}{ccc} 0 & 0 & -i \cr 
                   0 & 0 & 0 \cr
                   i & 0 & 0 \cr
\end{array}
\right), \\
\hspace{-.5cm}
&& 
\lambda_6= 
\left(
\begin{array}{ccc} 0 & 0 & 0 \cr 
                   0 & 0 & 1 \cr
                   0 & 1 & 0 \cr
\end{array}
\right),
\hspace{.1cm}
\lambda_7= 
\left(
\begin{array}{ccc} 0 & 0 & 0 \cr 
                   0 & 0 & -i \cr
                   0 & i & 0 \cr
\end{array}
\right), 
\lambda_8= 
\frac{1}{\sqrt{3}}
\left(
\begin{array}{ccc} 1 & 0 & 0 \cr 
                   0 & 1 & 0 \cr
                   0 & 0 & -2 \cr
\end{array}
\right). 
\nonumber
\ea}
For completeness we can also define $\lambda_0= {\rm diag}(1,1,1)$.

In the first line $\lambda_i$ ($i=1,2,3$)
are the SU(3) generalization
of the Pauli matrices $\tau_i$ ($i=1,2,3$).
In the second line the matrices $\lambda_4$ and $\lambda_5$ 
mix the quarks $u$ and $s$.
In the last line the matrices $\lambda_6$ and $\lambda_7$
mix $d$ and $s$.
The matrix $\lambda_8$ acts on all flavors ($u$, $d$ and $s$).

The neutral transitions ($\Delta I=0$ and  $\Delta S=0$) are 
associated with the operator
\ba
I_0 = \lambda_3.
\ea
The transitions that increase/decrease the isospin ($\Delta I=\pm 1$)
are defined by 
\ba
I_\pm = \frac{1}{2}( \lambda_1 \pm i \lambda_2). 
\ea
The transitions associated with  
$s \rightleftarrows u$ ($\Delta S =\pm 1$)
are represented by the operator
\ba
V_\pm = \frac{1}{2}( \lambda_4 \pm i \lambda_5). 
\ea

For the discussion of the quark electromagnetic current, 
it is convenient to define 
\ba
\lambda_s= 
\left(
\begin{array}{ccc} 0 & 0 & 0 \cr 
                   0 & 0 & 0 \cr
                   0 & 0 & -2 \cr
\end{array}
\right),
\ea
which can also be written as 
\ba
\lambda_s =  \sqrt{3} \lambda_8 - \tau_0,
\ea
where $\tau_0= {\rm diag}(1,1,0)$ is the SU(3)
generalization of the SU(2) unitary matrix.

\clearpage

\section{Numerical results for the axial-vector form factor $G_A$
  \label{app-GA}}

\setcounter{figure}{0}
\renewcommand\thefigure{\thesection\arabic{figure}}

We present here the numerical results
for the form factors $G_A$
for all transitions between octet baryon members
discussed in Secs.~\ref{sec-Vacuum}, \ref{sec-Medium1} 
and \ref{sec-Medium2}.
For a more clear comparison of magnitudes and shapes,
we present the results of equivalent functions
in the same column.

In the cases where the final result $G_A$ is negative,
we represent the function $-G_A$, for an easier comparison
with the other transitions.

In Fig.~\ref{figGA-p1X} we present the results for
the $|\Delta I| =1$ transitions, 
and in Figs.~\ref{figGA-p21X} and \ref{figGA-p22X} we present the
results for the $|\Delta S| =1$ transitions.

The axial-vector form
factors in the vacuum are on the right column [Sec.~\ref{sec-Vacuum}].
The contributions from the quark core are represented by the
dashed lines, and the contributions from the meson cloud by
the dashed-doted lines.
The final results correspond to the solid lines.
On the left column we present the final results of $G_A$
for the vacuum ($\rho= 0$)
and medium (densities $\rho= 0.5 \rho_0$ and $\rho= \rho_0$)
[Sec.~\ref{sec-Medium1}].

In Figs.~\ref{figGA-p30X} and \ref{figGA-p31X}
we represent the ratios $G_A^\ast(Q^2)/G_A(Q^2)$
on the right column, and
the ratios $G_A^\ast(Q^2)/G_A^\ast(0)$ on the left panel,
for the $|\Delta I| =1$ and $|\Delta S| =1$ cases, respectively
[Sec.~\ref{sec-Medium2}].

Concerning the $|\Delta I| =1$ transitions (Figs.~\ref{figGA-p1X} and \ref{figGA-p30X}),
we recall that the form factors $\Sigma^+ \to \Lambda$
and  $\Sigma^- \to \Lambda$ differ by a sign,
and that the form factors 
$\Sigma^- \to \Sigma^0$ and $\Sigma^0 \to \Sigma^+$ are equal.

As for the $|\Delta S| =1$ transitions (Figs.~\ref{figGA-p21X}, \ref{figGA-p22X} and \ref{figGA-p31X}),
the ratios for the transitions $\Sigma^- \to n$ and $\Sigma^0 \to p$
are the same.
The same is true also for the $\Xi^0 \to \Sigma^+$ and
$\Xi^- \to \Sigma^0$ transitions.

%%%%%%%%%%%%%%%%%%%%%%%%%%%%%%%%%%%%%%%%%%%%%%%%%%%%%%%%%%%%%%%%%%%%
%
%    FORM FACTORS RATIOS    GA*(Q2)/GA*(0)
%
%%%%%%%%%%%%%%%%%%%%%%%%%%%%%%%%%%%%%%%%%%%%%%%%%%%%%%%%%%%%%%%%%%%%

\begin{figure*}[h]   %%%%    Figure 3 (11)
%\vspace{.5cm}
\begin{center}
\mbox{
\includegraphics[width=2.9in]{GA-np} \hspace{1.cm}
\includegraphics[width=2.9in]{GA-np-M} }
\end{center}
 \vspace{.1cm}
\begin{center}
 \mbox{
\includegraphics[width=2.9in]{GA-SpL} \hspace{1.cm}
\includegraphics[width=2.9in]{GA-SpL-M} }
\end{center}
\vspace{.1cm}
\begin{center}
\mbox{
\includegraphics[width=2.9in]{GA-S0Sp} \hspace{1.cm}
\includegraphics[width=2.9in]{GA-S0Sp-M} }
\end{center}
\vspace{.1cm}
\begin{center}
\mbox{
\includegraphics[width=2.9in]{GA-XmX0B} \hspace{1.cm}
\includegraphics[width=2.9in]{GA-XmX0-M1} }
\end{center}
\caption{\footnotesize{Axial-vector form factor $G_A$ for $|\Delta I| =1$ transitions.
    {\bf Left panel:} results in vacuum (bare, meson cloud and total).
    {\bf Right panel:} total results for the medium $\rho=0.5 \rho_0$ and
    $\rho_0$ compared with vacuum ($\rho=0$).}}
\label{figGA-p1X}
\end{figure*}

\begin{figure*}[t]    %%%%    Figure 4 (12)
\vspace{.5cm}
\begin{center}
\mbox{
\includegraphics[width=2.9in]{GA-Lp} \hspace{1.cm}
\includegraphics[width=2.9in]{GA-Lp-M}}
\end{center}
 \vspace{.1cm}
\begin{center}
 \mbox{
\includegraphics[width=2.9in]{GA-Sn} \hspace{1.cm}
\includegraphics[width=2.9in]{GA-Sn-M} }
\end{center}
 \vspace{.1cm}
 \begin{center}
\mbox{
\includegraphics[width=2.9in]{GA-S0pB} \hspace{1.cm}
\includegraphics[width=2.9in]{GA-S0p-M1} }
\end{center}
\caption{\footnotesize{Axial-vector form factor $G_A$
for $|\Delta S| =1$ transitions (part 1).
    {\bf Left panel:} results in vacuum (bare, meson cloud and total).
    {\bf Right panel:} total results for the medium $\rho=0.5 \rho_0$ and
    $\rho_0$ compared with vacuum ($\rho=0$). }}
\label{figGA-p21X}
\end{figure*}

%%%%%%%%%%%%%%%%%%%%%%%%%%%%%%%%%%%%%%%%%%%%%%%%%%%%%%%%%%%%%%%%%%%%%%
%%%%%     FIGURES
%%%%%%%%%%%%%%%%%%%%%%%%%%%%%%%%%%%%%%%%%%%%%%%%%%%%%%%%%%%%%%%%%%%%%%

\begin{figure*}[t]    %%%%    Figure 5  (page 13)
%\vspace{.5cm}
\begin{center}
\mbox{
\includegraphics[width=2.9in]{GA-XmL} \hspace{1.cm}
\includegraphics[width=2.9in]{GA-XmL-M}}
\end{center}
 \vspace{.1cm}
\begin{center}
 \mbox{
\includegraphics[width=2.9in]{GA-XmS0} \hspace{1.cm}
\includegraphics[width=2.9in]{GA-XmS0-M} }
\end{center}
 \vspace{.1cm}
 \begin{center}
\mbox{
\includegraphics[width=2.9in]{GA-X0SpB} \hspace{1.cm}
\includegraphics[width=2.9in]{GA-X0Sp-M1} }
\end{center}
 \caption{\footnotesize{Axial-vector form factor $G_A$ for $|\Delta S| =1$ transitions (part 2).
    {\bf Left panel:} results in vacuum (bare, meson cloud and total).
    {\bf Right panel:} total results for the medium $\rho=0.5 \rho_0$ and
    $\rho_0$ compared with vacuum ($\rho=0$). }}
\label{figGA-p22X}
\end{figure*}

%%%%%%%%%%%%%%%%%%%%%%%%%%%%%%%%%%%%%%%%%%%%%%%%%%%%%%%%%%%%%%%%%%%%%

\begin{figure*}[t]   %%    Fig. 9  (17)
  %\vspace{.5cm}
\begin{center}
\mbox{
\includegraphics[width=2.9in]{GA-np-RT} \hspace{1.cm}
\includegraphics[width=2.9in]{GA-np-R} }
\end{center}
 \vspace{.1cm}
%\centerline{ \vspace{.1cm}}
\begin{center}
\mbox{
\includegraphics[width=2.9in]{GA-SpL-RT} \hspace{1.cm}
\includegraphics[width=2.9in]{GA-SpL-R} }
\end{center}
 \vspace{.1cm}
%\centerline{ \vspace{.1cm}}
\begin{center}
\mbox{
\includegraphics[width=2.9in]{GA-S0Sp-RT} \hspace{1.cm}
\includegraphics[width=2.9in]{GA-S0Sp-R}}
\end{center}
 \vspace{.1cm}
%\centerline{ \vspace{.1cm}}
\vspace{.1cm}
\begin{center}
\mbox{
\includegraphics[width=2.9in]{GA-XmX0-RTB} \hspace{1.cm}
\includegraphics[width=2.9in]{GA-XmX0-RB}}
\end{center}
\caption{\footnotesize{Axial-vector form factor $G_A$ for $|\Delta I| =1$ transitions.
    Ratios $G_A^\ast/G_A$ and $G_A^\ast(Q^2)/G_A^\ast(0)$.}}
\label{figGA-p30X}
\end{figure*}

\begin{figure*}[t]   %%   Fig.10   (19)
  %\vspace{.5cm}
\begin{center}
\mbox{
\includegraphics[width=2.9in]{GA-Lp-RT} \hspace{1.cm}
\includegraphics[width=2.9in]{GA-Lp-R}}
\end{center}
%\centerline{ \vspace{.05cm}}
 \vspace{.1cm}
\begin{center}
 \mbox{
\includegraphics[width=2.9in]{GA-SmN-RT} \hspace{1.cm}
\includegraphics[width=2.9in]{GA-SmN-R} }
\end{center}
 \vspace{.1cm}
%\centerline{ \vspace{.05cm}}
\begin{center}
\mbox{
\includegraphics[width=2.9in]{GA-XmL-RT} \hspace{1.cm}
\includegraphics[width=2.9in]{GA-XmL-R}}
\end{center}
\vspace{.1cm}
 \begin{center}
\mbox{
\includegraphics[width=2.9in]{GA-X0Sp-RTB} \hspace{1.cm}
\includegraphics[width=2.9in]{GA-X0Sp-RB} }  %GA-X0Sp-RB
\end{center}
\caption{\footnotesize{Axial-vector form factor $G_A$ for $|\Delta S| =1$ transitions.
    Ratios $G_A^\ast/G_A$ and $G_A^\ast(Q^2)/G_A^\ast(0)$. 
     The ratios are the same for  $\Xi^- \to \Sigma^0$ and $\Xi^0 \to \Sigma^+$.
    The ratios are the same for $\Sigma^- \to n$ and $\Sigma^0 \to p$.}  }
\label{figGA-p31X}
\end{figure*}

\section{Numerical results for the induced pseudoscalar form factor $G_P$
  \label{app-GP}}

\setcounter{figure}{0}
\renewcommand\thefigure{\thesection\arabic{figure}}

We present here the numerical results
for the form factors $G_P$
for all transitions between octet baryon members
discussed in Secs.~\ref{sec-Vacuum}, \ref{sec-Medium1} 
and \ref{sec-Medium2}.
For a more clear comparison of magnitudes and shapes
we present the results of equivalent functions
in the same column.

As for the function $G_A$, in the cases where the final result
$G_P$ is negative,
we represent the function $-G_P$.

In Fig.~\ref{figGP-p1X} we present the results for
the $|\Delta I| =1$ transitions
and in Figs.~\ref{figGP-p21X} and  \ref{figGP-p22X} we present the
results for the $|\Delta S| =1$ transitions.

On the right column we present the induced pseudoscalar form
factors in the vacuum [Sec.~\ref{sec-Vacuum}].
In the figures, we discriminate the result for the bare contribution,
the combination of the bare and pseudoscalar meson pole contribution,
and the sum of all the contributions (Bare + Pole + Meson Cloud).
On the left column we present the final results
for the vacuum ($\rho= 0$)
and medium (densities $\rho= 0.5 \rho_0$ and $\rho= \rho_0$)
[Sec.~\ref{sec-Medium1}].

In Figs.~\ref{figGP-p30X} and \ref{figGP-p31X}, we represent the ratios
$G_P^\ast(Q^2)/G_P(Q^2)$
on the right column, and
and the ratio $G_P^\ast(Q^2)/G_P(Q_p^2)$
where $Q_p^2= 0.012$ GeV$^2$, on the left column,
for the $|\Delta I| =1$ and $|\Delta S| =1$ cases, respectively
[Sec.~\ref{sec-Medium2}].

As in the case of the form factor $G_A$, there
are some relations between the theoretical calculations
associated with different channels.
For the $|\Delta I| =1$ transitions (Figs.~\ref{figGP-p1X} and \ref{figGP-p30X}),
we recall that the form factors $\Sigma^+ \to \Lambda$
and  $\Sigma^- \to \Lambda$ differ by a sign,
and that the form factors 
$\Sigma^- \to \Sigma^0$ and $\Sigma^0 \to \Sigma^+$ are the same.

As for the $|\Delta S| =1$ transitions (Figs.~\ref{figGP-p21X},  \ref{figGP-p22X}
and \ref{figGP-p31X}), the ratios for the transitions $\Sigma^- \to n$ and $\Sigma^0 \to p$
are the same.
The same property is also valid for the $\Xi^0 \to \Sigma^+$ and
$\Xi^- \to \Sigma^0$ transitions.

%%%%%%%%%%%%%%%%%%%%%%%%%%%%%%%%%%%%%%%%%%%%%%%%%%%%%%%%%%%%%%%%%%%%%%%%%%%%%%%
%%%%%%%%%%%%%%%%%%%%%%%%%%%%%%%%%%%%%%%%%%%%%%%%%%%%%%%%%%%%%%%%%%%%%%%%%%%%%%%
%%%%%%%%%%%%%%%%%%%%%%%%%%%%%%%%%%%%%%%%%%%%%%%%%%%%%%%%%%%%%%%%%%%%%%%%%%%%%%%

\begin{figure*}[t]    %%   Fig. 6  (14)
  %\vspace{.5cm}
\begin{center}
\mbox{
\includegraphics[width=2.9in]{GP-np} \hspace{1.cm}
\includegraphics[width=2.9in]{GP-np-M} }
\end{center}
 \vspace{.1cm}
\begin{center}
 \mbox{
\includegraphics[width=2.9in]{GP-SpL} \hspace{1.cm}
\includegraphics[width=2.9in]{GP-SpL-M} }
\end{center}
\vspace{.1cm}
\begin{center}
 \mbox{
\includegraphics[width=2.9in]{GP-S0Sp} \hspace{1.cm}
\includegraphics[width=2.9in]{GP-S0Sp-M} }
\end{center}
\vspace{.1cm}
\begin{center}
 \mbox{
\includegraphics[width=2.9in]{GP-XmX0} \hspace{1.cm}
\includegraphics[width=2.9in]{GP-XmX0-M} }
\end{center}
\caption{\footnotesize{Induced pseudoscalar form factor $G_P$ for $|\Delta I| =1$ transitions.
    {\bf Left panel:} results for bare, bare plus meson cloud and total.
    {\bf Right panel:} total results for the medium $\rho=0.5 \rho_0$ and
    $\rho_0$ compared with vacuum ($\rho=0$).}}
\label{figGP-p1X}
\end{figure*}

\begin{figure*}[t]    %%%%   Figure 7  (page 15)
  %\vspace{.5cm}
\begin{center}
\mbox{
\includegraphics[width=2.9in]{GP-Lp} \hspace{1.cm}
\includegraphics[width=2.9in]{GP-Lp-M}}
\end{center}
 \vspace{.18cm}
\begin{center}
 \mbox{
\includegraphics[width=2.9in]{GP-SmN} \hspace{1.cm}
\includegraphics[width=2.9in]{GP-SmN-M} }
\end{center}
 \vspace{.1cm}
 \begin{center}
\mbox{
\includegraphics[width=2.9in]{GP-S0pB} \hspace{1.cm}
\includegraphics[width=2.9in]{GP-S0p-M1} }
\end{center}
 \caption{\footnotesize{Induced pseudoscalar form factor $G_P$
     for $|\Delta S| =1$ transitions
(part 1).
     {\bf Left panel:} results for bare, bare plus meson cloud and total.
{\bf Right panel:} total results for the medium $\rho=0.5 \rho_0$ and
$\rho_0$ compared with vacuum ($\rho=0$).
 }}
\label{figGP-p21X}
\end{figure*}

\begin{figure*}[t]     %%%%   Figure 8  (page 16)
  %\vspace{.5cm}
\begin{center}
\mbox{
\includegraphics[width=2.9in]{GP-XmL} \hspace{1.cm}
\includegraphics[width=2.9in]{GP-XmL-M}}
\end{center}
 \vspace{.1cm}
\begin{center}
 \mbox{
\includegraphics[width=2.9in]{GP-XmS0} \hspace{1.cm}
\includegraphics[width=2.9in]{GP-XmS0-M} }
\end{center}
 \vspace{.18cm}
 \begin{center}
\mbox{
\includegraphics[width=2.9in]{GP-X0Sp} \hspace{1.cm}
\includegraphics[width=2.9in]{GP-X0Sp-M} }
\end{center}
 \caption{\footnotesize{Induced pseudoscalar form factor $G_P$
 for $|\Delta S| =1$ transitions (part 2).
     {\bf Left panel:} results for bare, bare plus meson cloud and total.
     {\bf Right panel:} total results for the medium $\rho=0.5 \rho_0$ and
$\rho_0$ compared with vacuum ($\rho=0$).
    }}
\label{figGP-p22X}
\end{figure*}

\begin{figure*}[t]      %%  Fig. 11  (21)
  %\vspace{.5cm}   %%  vspace can be avoided
\begin{center}
\mbox{
\includegraphics[width=2.8in]{GP-np-RT} \hspace{1.cm}
\includegraphics[width=2.9in]{GP-np-R} }
\end{center}
\vspace{.18cm}
\begin{center}
\mbox{
\includegraphics[width=2.9in]{GP-SpL-RT} \hspace{1.cm}
\includegraphics[width=2.9in]{GP-SpL-R} }
\end{center}
 \vspace{.13cm}
\begin{center}
\mbox{
\includegraphics[width=2.8in]{GP-S0Sp-RT} \hspace{1.cm}
\includegraphics[width=2.9in]{GP-S0Sp-R}}
\end{center}
\vspace{.13cm}
\begin{center}
\mbox{
\includegraphics[width=2.8in]{GP-XmX0-RTB} \hspace{1.cm}
\includegraphics[width=2.9in]{GP-XmX0-RB}}
\end{center}
\caption{\footnotesize{Induced pseudoscalar form factor
    $G_P$ for $|\Delta I| =1$ transitions.
    Ratios $G_P^\ast/G_P$ and $G_P^\ast(Q^2)/G_P^\ast(Q_p^2)$.  }}
\label{figGP-p30X}
\end{figure*}

\begin{figure*}[t]    %%  Fig. 12  (23)
  %\vspace{.5cm}
\begin{center}
\mbox{
\includegraphics[width=2.8in]{GP-Lp-RT} \hspace{1.cm}
\includegraphics[width=2.9in]{GP-Lp-R}}
\end{center}
 \vspace{.15cm}
\begin{center}
 \mbox{
\includegraphics[width=2.8in]{GP-SmN-RT} \hspace{1.cm}
\includegraphics[width=2.9in]{GP-SmN-R} }
\end{center}
\vspace{.15cm}
\begin{center}
\mbox{
\includegraphics[width=2.8in]{GP-XmL-RT} \hspace{1.cm}
\includegraphics[width=2.9in]{GP-XmL-R}}
\end{center}
\vspace{.15cm}
 \begin{center}
\mbox{
\includegraphics[width=2.9in]{GP-X0Sp-RTB} \hspace{1.cm}
\includegraphics[width=2.9in]{GP-X0Sp-RB} }
\end{center}
\caption{\footnotesize{Induced pseudoscalar form factor $G_P$ for $|\Delta S| =1$ transitions.
    Ratios $G_P^\ast/G_P$ and $G_P^\ast(Q^2)/G_P^\ast(Q_p^2)$. 
    The ratios are the same for  $\Xi^- \to \Sigma^0$ and $\Xi^0 \to \Sigma^+$.
    The ratios are the same for $\Sigma^- \to n$ and $\Sigma^0 \to p$.
}}
\label{figGP-p31X}
\end{figure*}

\end{document}